\documentclass[10pt]{article}

\usepackage{enumerate}
\usepackage{graphicx}
\usepackage{float}
\usepackage{caption}
\usepackage{subcaption}
\usepackage{amsfonts}
\usepackage{amssymb}
\usepackage{amsmath}
\usepackage{amsthm}
\usepackage{amsfonts}
\usepackage{siunitx}
\usepackage{gensymb}
\usepackage{textcomp}
\usepackage{tabu,multirow}
\usepackage{tabto}
\usepackage{stackengine}

\newcommand\xrowht[2][0]{\addstackgap[.5\dimexpr#2\relax]{\vphantom{#1}}}

\newcommand{\R}{\mathbb R}
\newcommand{\N}{\mathbb N}

\usepackage{xcolor,todonotes}

\begin{document}

\title{Quantification of Bore Path Uncertainty in Borehole Heat Exchanger Arrays}
\author{Philipp Steinbach$^{1}$, Daniel Otto Schulte$^{2}$, Bastian Welsch$^{2}$\\ Ingo Sass$^{2}$, Jens Lang$^{1}$\footnote{corresponding author}\\
\small \it steinbach@mathematik.tu-darmstadt.de, daniel.schulte@gast.tu-darmstadt.de\\
\small \it welsch@geo.tu-darmstadt.de, sass@geo.tu-darmstadt.de, lang@mathematik.tu-darmstadt.de\\
\small \it 1: Technische Universit\"at Darmstadt, Dolivostra{\ss}e 15, 64293 Darmstadt, Germany\\
\small \it 2: Technische Universit\"at Darmstadt, Schnittspahnstraße 9, 64287 Darmstadt, Germany\\}

\maketitle

\begin{abstract}
Borehole heat exchanger arrays have become a common implement for the utilization of thermal energy in the soil. Building these facilities is expensive, especially the drilling of boreholes, into which closed-pipe heat exchangers are inserted. Therefore, cost-reducing drilling methods are common practice, which can produce inaccuracies of varying degree. This brings into question how much these inaccuracies could potentially affect the performance of a planned system. In the presented case study, an uncertainty quantification for seasonally operated borehole heat exchanger arrays is performed to analyze the bore paths' deviations impact. We introduce an adaptive, anisotropic stochastic collocation method, known as the generalized Smolyak algorithm, which was previously unused in this context and apply it to a numerical model of the borehole heat exchanger array. Our results show that the borehole heat exchanger array performance is surprisingly reliable even with potentially severe implementation errors during their construction. This, coupled with the potential uses of the presented method in similar applications gives planners and investors valuable information regarding the viability of borehole heat exchanger arrays in the face of uncertainty. With this paper, we hope to provide a powerful statistical tool to the field of geothermal energy, in which uncertainty quantification methods are still rarely used at this point. The discussed case study represents a jumping-off point for further investigations on the effects of uncertainty on borehole heat exchanger arrays and borehole thermal energy storage systems.
\end{abstract}

\noindent {\bf Keywords}: uncertainty quantification; stochastic collocation; Smolyak sparse grids; generalized Smolyak algorithm; adaptivity; simulation; finite element method; borehole heat exchanger arrays; borehole path; vertical ground-source heat pump system;
\newline\newline
\noindent {\bf AMS}: 65C20; 65C60; 86-08;

\newpage

\section{Introduction}

Ground-source heat pump systems are a key technology to decarbonize the heat supply of buildings by replacing fossil fuels with geothermal energy \cite{MUSTAFAOMER2008reviewgshp}. They have several advantages over air-source heat pump systems, such as higher efficiency, lower life-cycle costs, lower environmental impact, higher reliability, and other practical benefits \cite{Wu2009_ashp_vs_gshp}.

Systems that use closed-loop vertical pipes (i.e. U-pipes, double U-pipes or coaxial pipes) to extract heat from the ground, called borehole heat exchangers (BHE), are advantageous over other closed-loop systems. They require less total pipe length, lower pumping energy, and the least amount of surface ground area \cite{MUSTAFAOMER2008reviewgshp}. Moreover, they also achieve higher efficiencies because soil temperatures at larger depths are not subject to seasonal fluctuations. Compared to open-loop systems (i.e. groundwater well doublets), they are much less site dependent.

\begin{figure}[t!]
        \centering
        \includegraphics[width=0.75\textwidth]{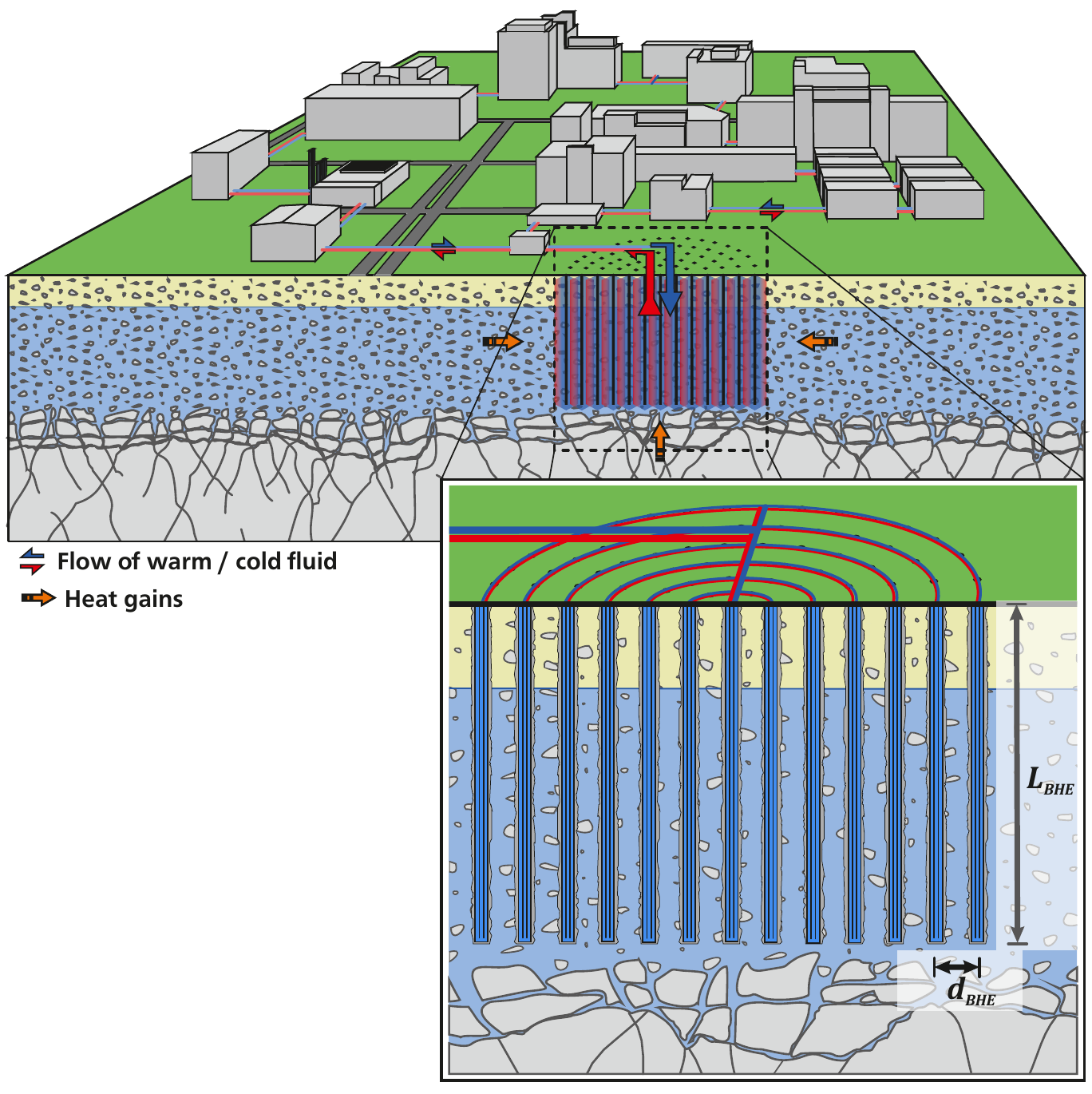}
        \caption{A BHE array operated in a district heating application. The individual BHEs are arranged in a circular pattern. Heat energy is extracted from the soil by pumping cold water (blue arrows) through the BHE array and then distributed to the surrounding buildings via the heated water (red arrows).}
        \label{fig:_seasonal_BHEA_facility}
\end{figure}

BHE pipes are installed in wellbores with a depth of typically 45-75 m for residential buildings and over 150 m for larger industrial uses \cite{SELF2013341}. To ensure a thermal connection to the subsurface and to prevent groundwater flow along the BHE axis, the annular gap between such a pipe system and the borehole wall is usually filled with a cement-based grouting material \cite{YANG2010verticalghe,Anbergen2014ftw}. A heat transfer fluid circulates through the BHE pipes and extracts heat from the natural ground. Heat pumps are usually necessary to shift the temperature of the heat provided to the required supply temperature of the heating system.
In district heating applications, systems consisting of several BHEs in a dense array (e.g., \cite{naicker2020bhearray}) have become common because they are less expensive than individual BHE. A typical example for a BHE array in a heat extraction application is sketched in Figure \ref{fig:_seasonal_BHEA_facility}.

\begin{figure}[t!]
        \centering
        \begin{subfigure}[b]{0.42\textwidth}
            \centering
            \includegraphics[width=\textwidth]{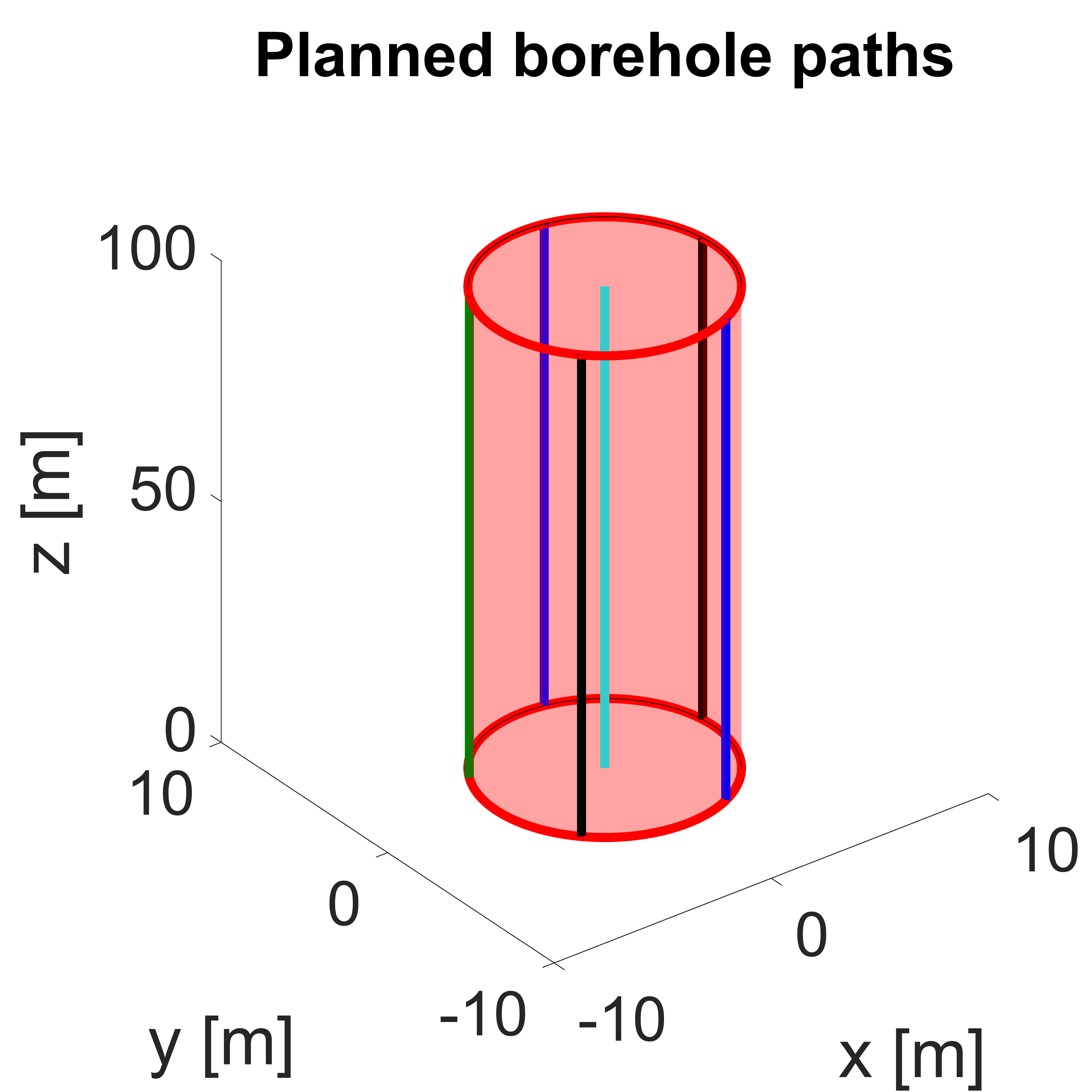}
            %\caption[Network2]%
            %{{\small Network 1}}
            %\label{fig:mean and std of net14}
        \end{subfigure}
        %\hfill
        \begin{subfigure}[b]{0.42\textwidth}
            \centering
            \includegraphics[width=\textwidth]{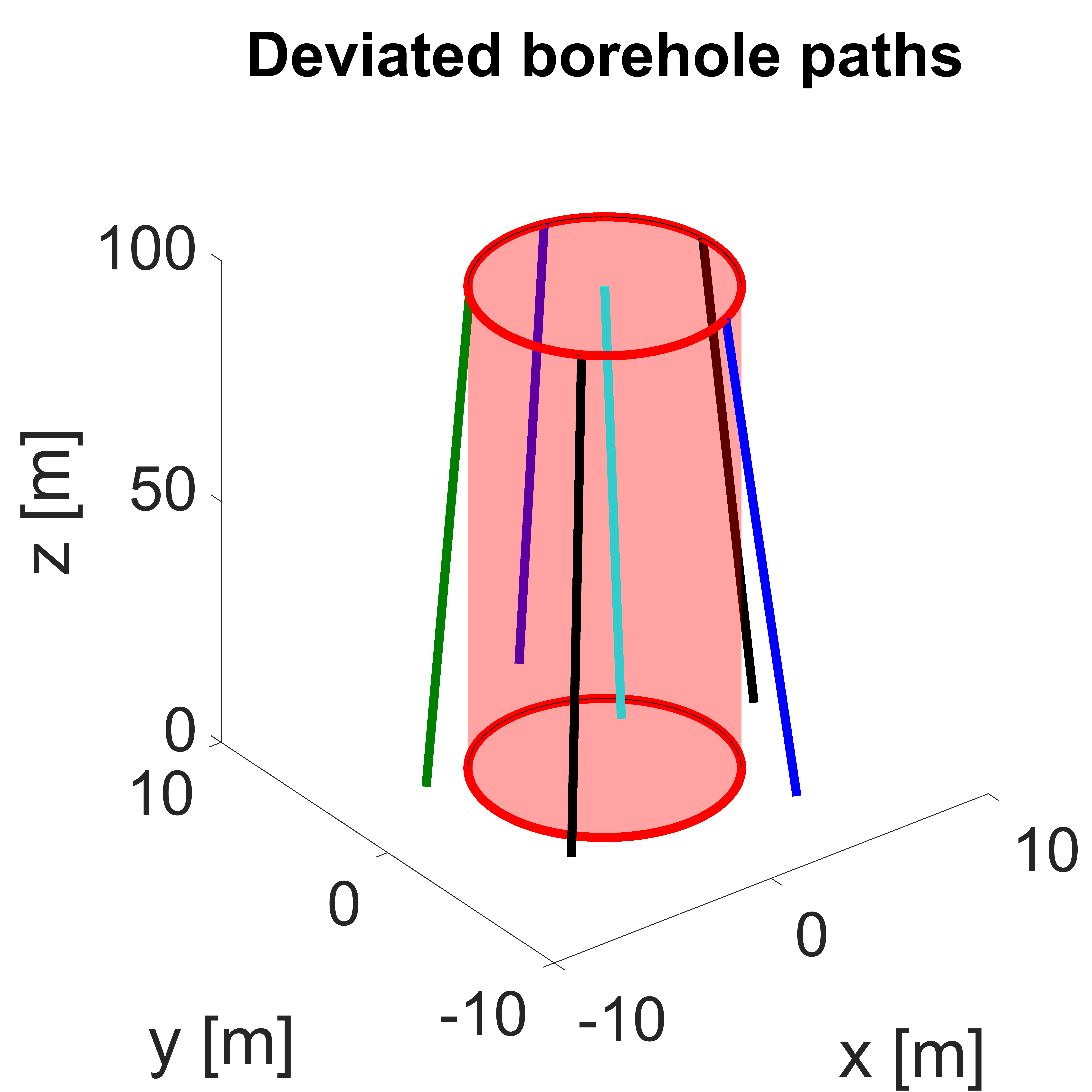}
            %\caption[]%
            %{{\small Network 2}}
            %\label{fig:mean and std of net24}
        \end{subfigure}
        \caption{A planned layout of boreholes (left) and a possible implementation with random deviations due to drilling inaccuracies (right).}
        \label{fig:deviations}
    \end{figure}

The drilling of boreholes is one of the most expensive steps during construction of these facilities \cite{blum2011techno}. There are multiple drilling techniques of varying cost and accuracy, although those of the highest accuracy are typically unaffordable for the construction of BHE arrays. Yet, even among the cheaper solutions there is always an incentive to cut costs, possibly at the expense to drilling accuracy. In reality, basically all BHE arrays are constructed with some amount of implementation errors, namely borehole paths deviating unpredictably from their planned vertical course, as sketched in Figure \ref{fig:deviations}.

Simulations and real world applications have shown, that the spacing of BHEs has a strong influence on the performance of BHE arrays \cite{GULTEKINspacing,QIAN2014spacing,MARCOTTE2010spacing}.
The closer the spacing gets, the more the individual BHE cannibalize each other leading to an overall inefficient system. However, due to space limitations in urban areas, systems with too large a BHE spacing become uneconomical, as well. As a result, a compromise between these two impairments must be found when designing BHE arrays.

However, in the vast majority of cases, the design of such systems is based on the assumption that the boreholes are exactly vertical. Probable, drilling-related and more or less random borehole deviations are disregarded. It is not yet possible to make a statistically reliable statement about the influence of bore path uncertainty on the performance of a BHE array. For this reason, this study provides an uncertainty quantification of the effects of variations in drilling trajectories on the performance of a BHE array.

Uncertainty quantification denotes the quantitative characterization of uncertainties, e.g. in models, which cannot be described deterministically due to a lack of exact information. Because of its broad definition, uncertainty quantification can be applied in almost any context, where academic models and real world applications meet. It is widely discussed and used and comprises a plethora of methods and applications \cite{ghanem2017handbook}.

The particular case study presented in this paper deals with a model in the form of a system of partial differential equations with up to $18$ random variables. There are several methods to approach this problem, including the stochastic finite element method \cite{stefanou2009stochastic}, Monte Carlo simulation \cite{teckentrup2013multilevel} and stochastic collocation \cite{xiu2005high}. For several reasons, the collocation approach was chosen to address uncertainties in BHE arrays: a deterministic finite element solver was available, which had been developed for the simulation of BHE arrays. This should be further used without changes to the code. Together with comparatively high stochastic dimension, this ruled out the stochastic finite element method and other non-sampling approaches. Moreover, the stochastic dimension of the problem is still moderate enough for stochastic collocation and a certain regularity of the solution was considered likely. Therefore, stochastic collocation was expected to be much more efficient during sampling than Monte Carlo. Additionally, some amount of anisotropy in the stochastic dimensions was assumed, which also favors the adaptive approach of the stochastic collocation method.

Modeling and quantification of uncertainties has been common practice in the field of geology for some time. One of the main applications here is subsurface modeling, where it is difficult and expensive to obtain accurate measurements of soil and rock properties, formation boundaries, and hydrogeologic data over large areas. Various methods to quantify and minimize uncertainty have been discussed in this context \cite{caers2005petroleum, chiles2009geostatistics} and have led to the development of widespread tools such as the software library GSLIB \cite{deutsch1998gslib}.

In the field of geothermal energy, the structure and composition of the subsurface is of course crucial and because of the natural overlap in the subject matter, the concepts and techniques apply here as well. There are several studies, for example, \cite{shetty2018numerical, vogt2010quantifying, vogt2010reducing, SCHULTE2020101792}, where the viability of geothermal reservoirs in the presence of uncertainties in the subsurface is discussed. Similarly, stochastic modeling of the subsurface is performed by \cite{asrizal2006uncertainty, yin2020automated}. Moreover, Lukawski et al. \cite{lukawski2016uncertainty} used uncertainty quantification methods to estimate geothermal construction costs.

Even though many concepts and techniques should extend to geothermal applications, uncertainty quantification is still new in this context and, as pointed out in \cite{vogt2010quantifying, witter2014uncertainty}, are rarely used. Apart from the sophisticated subsurface modelling, the utilized techniques are mostly simple. In parameter studies, sometimes just comparisons between discrete realizations \cite{limberger2018geothermal} are carried out, without accounting for a distribution on the parameter space. When accounting for distributions, usually Monte Carlo methods are used \cite{lukawski2016uncertainty, vogt2010quantifying, yin2020automated}. They are known to be easily integrated and they work quite well, as long as enough samples are available. However, there is a lot of untapped potential for improved efficiency and additional statistical information in the application of more advanced techniques. This study aims to contribute to more awareness in the choice of the uncertainty quantification method. It applies the stochastic collocation method, previously unused in this context, to the so far unexplored problem of borehole deviations in BHE arrays.

\section{Methodology}

Our goal is to study the effects of randomly occurring deviations in the layout geometry of a BHE array on its performance. The performance of an array will be measured by a characteristic number (quantity of interest) such as a storage coefficient in case of a storage/extraction scenario or an averaged inlet-outlet-temperature in case of a pure extraction scenario. The quantity of interest is obtained by simulating the array via a system of partial differential equations (PDE), which will be solved using the finite element method. We start out by introducing a deterministic BHE array model, which will then be extended to account for random deviations in the geometry. With the introduced uncertainty, the quantity of interest measuring performance is itself a random variable with stochastic properties, such as an expected value and variance. These properties can be obtained and studied by evaluating the stochastic setting using a stochastic collocation method. In this section, this approach will be discussed step by step in detail.

\subsection{Deterministic model of a BHE array as a system of partial differential equations}

We introduce a deterministic model of a BHE array as a system of PDEs, which will be solved using the finite element method solver KARDOS \cite{erdmann2006kardos}.  We implement an improved version of BASIMO \cite{SCHULTE2016210} in KARDOS, which couples the FEM problem based on the PDE with a thermal resistance and capacity model (TRCM) for BHEs \cite{bauer2011thermal}. BASIMO allows for the simulation of a BHE array being operated in various applications and was intended as a continuation of the work in \cite{schulte2016}, which resulted in the simulation tool BASIMO.

As the baseline model, we will be using a system of PDEs modeling time-dependent heat transfer in a saturated porous medium with groundwater flow, based on a model in \cite{lang2001heatmass} which also considers salt concentration. After reduction to the relevant components, we acquire a pressure-temperature model in the form of the coupled parabolic system
\begin{equation}\label{base_pde}
	\begin{aligned}
		n \rho \left(\beta \frac{\partial p}{\partial t} +\alpha \frac{\partial T}{\partial t} \right) + \nabla \cdot (\rho  \textbf{q}) &~ = ~  0 ~,		\\
		\left(n c \rho + (1 - n) \rho^s c^s 		\right )  \frac{\partial T}{\partial t} + \rho c \textbf{q} \cdot \nabla T + \nabla \cdot J_T & ~ = ~ F(x,y,z,t,T) ~,
	\end{aligned}
\end{equation}
with the solution functions $p = p(x,y,z,t)$ for the pressure and $T =T(x,y,z,t)$ for the temperature, depending on a three-dimensional region $\Omega$ and the time $t \in [0,\tau]$. $F$ describes heat sources and sinks, which in this practical example are the BHEs. The equations contain the Darcy velocity
\begin{align}
	\textbf{q}~ := ~ - \frac{K}{\mu}\left (\nabla p ~ - ~ \rho  \textbf{g} \right)
\end{align}
with the permeability tensor $K$ of the porous medium and the acceleration of gravity vector $\textbf{g} = (0,0,-g)^\intercal$ with $g=9.81\frac{m}{s^2}$. This also accounts for the hydrostatic pressure in the soil. Should one only be interested in modeling the pressure head, set $\textbf{g} = (0,0,0)^\intercal$. We assume $K$ to be a diagonal matrix, $K=\text{diag}(k)$. The heat flux vector $J_T$ is defined as
\begin{align}
	J_T ~ := ~ - \left((\kappa +\lambda_T |\textbf{q}|) I + \frac{\lambda_L - \lambda_T}{|\textbf{q}|}\textbf{q} \textbf{q}^T\right) \nabla T
\end{align}
with $|\textbf{q}| = \sqrt{\textbf{q}^T \textbf{q}}$. Table \ref{parameters_PDE} lists all parameters used in the PDE.

\begin{table}[t!]
		\centering
		\begin{tabu}{ |l|l|l|}
		\hline
		\multicolumn{3}{|c|}{model parameters} \\
		\hline
		symbol & parameter name & unit \\
		\hline
		$\alpha$ & temperature coefficient, fluid & $K^{-1}$ \\
		$\beta$ & compressibility, fluid & $m ~ s^2 ~ (kg)^{-1}$ \\
		$c$ & specific heat capacity, fluid & $ J ~ (kg)^{-1} ~ K^{-1}$  \\
		$\rho$ & density, fluid & $ (kg)~ m^{-3}$  \\
		$\mu$ & viscosity, fluid& $(kg)~ m^{-1} ~ s^{-1}$ \\
		$ n $ & porosity &  \\
		$ k $ & permeability & $m^2$ \\
		$\lambda_L$ & longitudinal heat conductivity & $ J ~ m^{-2} ~ K^{-1}$ \\
		$\lambda_T$ & transverse heat conductivity & $ J ~ m^{-2} ~ K^{-1}$  \\
		$\rho^s c^s$ & vol. heat capacity, solid/rock & $J ~  m^{-3} ~ K^{-1}$ \\
		$\kappa$ & thermal conductivity & $W ~ m^{-1} ~ K^{-1}$ \\
		\hline
		\end{tabu}
		\caption{Parameters used in the PDE}
		\label{parameters_PDE}
\end{table}

For the example discussed in this paper, we assume a homogeneous region $\Omega$, however all parameters may be considered space- and/or time-dependent.

\subsubsection{Modeling of the BHE as heat sources/sinks}
\label{sec:RightSide}

A fully discretized model of a BHE is computationally costly as the resulting FEM mesh requires a very large number of nodes to account for the greatly differing scales of the simulated objects. To avoid this, we model a BHE as a 1D line source embedded into the 3D FEM mesh. For this, we will make use of the thermal resistance and capacity model (TRCM) proposed in \cite{bauer2011thermal} based on \cite{eskilson1988simulation}, which simulates the internal state of a BHE.

The TRCM describes the internal state of the BHE via the temperature of its descending (inlet) and ascending (outlet) water $T_{in} = T_{in} (z,t)$ and $T_{out}= T_{out}(z,t)$ as two coupled functions depending on the depth $z$ (from $z=0$ to the length of the BHE $z=L$) and the time $t$. They are
\begin{align}
	\begin{aligned}
	%\label{T_in_einfach}
	T_{in}(z,t) = T_{in}(0,t)f_1(z) + T_{out}(0,t)f_2(z) + \int_0^z T_b(\zeta,t)f_4(z-\zeta) d\zeta ~,  \\
	%\label{T_out_einfach}
	T_{out}(z,t) = -T_{in}(0,t)f_2(z) + T_{out}(0,t)f_3(z) - \int_0^z T_b(\zeta,t)f_5(z-\zeta) d\zeta ~.
	\end{aligned}
\end{align}
Here, $T_b = T_b(z,t)$ describes the temperature of the borehole wall, which in our context is the temperature along the line, at which the BHE line source will be embedded into $\Omega$. The auxiliary functions $f_1,...,f_5$ are given by
\begin{align}
	\begin{aligned}
	&f_1(z) = e^{\beta z} [\cosh(\gamma z) - \delta\sinh(\gamma z)] ~, 												\\
	&f_2(z) = e^{\beta z} \frac{\beta_{12}}{\gamma}\sinh(\gamma z) ~, 												\\
	&f_3(z) = e^{\beta z} [\cosh(\gamma z) + \delta\sinh(\gamma z)]	~, 												\\
	&f_4(z) = e^{\beta z} \left[\beta_1\cosh(\gamma z) - \left(\delta \beta_1 +\frac{\beta_2\beta_{12}}{\gamma} \right) \sinh(\gamma z)\right] ~, 	\\
	&f_5(z) = e^{\beta z} \left[\beta_2\cosh(\gamma z) + \left(\delta \beta_2 +\frac{\beta_1\beta_{12}}{\gamma} \right) \sinh(\gamma z)\right] ~,
	\end{aligned}
\end{align}
with $\alpha$, $\beta$, $\beta_1$, $\beta_2$, $\beta_{12}$, $\gamma$ and $\delta$ being meta parameters of the material parameters, geometric measurements and operation parameters. For a detailed description, see \cite{eskilson1988simulation}. Both $T_{in} (z,t)$ and $T_{out}(z,t)$ can be transformed to an explicit expression without referencing each other, allowing us to the formulate a heat source/sink function
\begin{align}
	F_{U}(t, z) = \frac{T_{in}(z,t)-T_b(z,t)}{R} + \frac{T_{out}(z,t)-T_b(z,t)}{R}
\end{align}
based on the temperature gradient of the carrier fluid and the borehole wall. $R$ describes a thermal resistance calculated in the TRCM. $F_U$ relates to a U-type BHE, but analogue functions can be derived for double U- and coaxial-type BHEs. Multiple $F_U$-terms, or their counterparts for other BHE-types, form the $F$ in (\ref{base_pde}) via a coupling-function depending on the number of BHEs and the geometric layout of the array.

This definition of the heat source/sink means that the FEM model and the TRCM are coupled reciprocally. For every point in time, the FEM model feeds the TRCM the current borehole temperature and in turn, the TRCM provides a source/sink term to the FEM model.

The coupled FEM solution was implemented and validated in \cite{steinbach2018} with multiple benchmarks calculated with FEFLOW \cite{diersch2013feflow}.

\subsection{Introducing uncertainty into the model}
\label{modeling_uncertainty}

(\ref{base_pde}) can be written as an abstract PDE with partial differential operator $A$, boundary condition operator $B$ and the solution function $v=(p, T)$ with initial condition $v_0$. The new system equation reads
\begin{equation}
\label{abstract_pde}
	\begin{aligned}
		A(v, x, t) &=  f(v,x,t)  ~~~ \text{in } \Omega \times [0,\tau],	  \\
		B(v, x, t) &=  g(v,x,t) ~~~ \text{in } \partial\Omega \times [0,\tau],	  \\
		v(x,0) &= v_0(x) ~~~ \text{in } \Omega.
	\end{aligned}
\end{equation}

We now introduce a finite set of stochastically independent random variables $\mathcal{X}_1,~ \mathcal{X}_2,~ \dots~ \mathcal{X}_M$ with probability density functions (pdf) $\rho_1,~\rho_2,~\dots~\rho_M$ on the the joint support $\Gamma = \Gamma_1~\times~\Gamma_2~\times~\cdots~\times~\Gamma_M$ and extend (\ref{abstract_pde}) to a parametrized system of PDEs
\begin{equation}
\label{abstract_pde_extended}
	\begin{aligned}
		A(v, x, t, y) &=  f(v,x,t,y)  ~~~ \text{in } \Omega \times [0,\tau] \times \Gamma,  \\
		B(v, x, t, y) &=  g(v,x,t,y) ~~~ \text{in } \partial\Omega \times [0,\tau] \times \Gamma,	\\
		v(x,0, y) &= v_0(x,y) ~~~ \text{in } \Omega \times \Gamma.
	\end{aligned}
\end{equation}
with $y \in \Gamma$  denoting a point in the stochastic parameter space of finite dimension N
(finite noise assumption). Formally, the PDE and its solution $v = v(x,t,y)$ now also depend on the support of the random variables. This allows us to study the effects of uncertainty on various components of the PDE in an analytical setting. As a simple example, we can assume one of the soil parameters in the PDE such as the heat conductivity $\kappa$ to be uncertain. Instead of a deterministic value for $\kappa$, we introduce a random variable $\mathcal{X}_\kappa$ with a probability density function $\rho_\kappa$ on the support $\Gamma_\kappa = [\kappa_{\text{min}},\kappa_{\text{max}}]$. The PDE is then considered on the entire range $\kappa \in \Gamma_\kappa$  and we can investigate the (now stochastic) solution $v$ with respect to $\rho_\kappa$.

Theoretically, we can consider $v = v(x,t,y)$ as fully space-, time- and uncertainty-dependent. However, to make the solution more manageable, we break it down to a quantity of interest $u= u(y) = \phi (v(x,t,y))$, which is no longer space- and time-dependent. Here, $\phi$ is an abstract function, that maps the fully dependent solution $v$ onto the considered quantity of interest $u$. For our practical application in the context of BHE arrays, the quantity of interest will be an inlet-outlet-temperature averaged for all BHEs in an array and also the simulated time of operation. The averaged inlet-outlet-temperature is a real number that we can compare for each realization $y \in \Gamma$, which only depends implicitly on $\Omega$ and the simulated time frame $[0,\tau]$. For detailed application of this, see section \ref{sec:example}. Other quantities of interest may be considered as well.

As motivated earlier, we are particularly interested in the effects of randomly occurring deviations in the BHE array geometry, which we will model in the following section.

\subsubsection{Modeling of uncertainty in the BHE array geometry}
\label{sec:modeling}

Using the described FEM model, we simulate a cuboid patch of soil bordering the earth's surface at the top. A point in space will be addressed by $(x,y,z)$, with $(x,y)$ identifying the horizontal position and the depth $z$. In this patch of soil, we place multiple BHEs which are modeled as dynamic line heat sources/sinks depending on the temperature of the soil and the temperature of the carrier fluid running through the BHEs. We assume that for the original, planned facility, all BHEs are placed at the top of the patch of soil, bordering the surface. They each form a line perpendicular to the surface going into soil to a certain depth.

Next, we introduce random variables, which will govern azimuth and inclination of each BHE's bore path. For the upcoming uncertainty quantification, we want to study the effects of linearly deviating borehole paths (possible bore path curvature is not considered). This can be modeled using two random variables for each BHE, for which we want to allow deviations to occur.

\begin{figure}[htbp]
        \centering
        \begin{subfigure}[b]{0.35\textwidth}
            \centering
            \includegraphics[width=\textwidth]{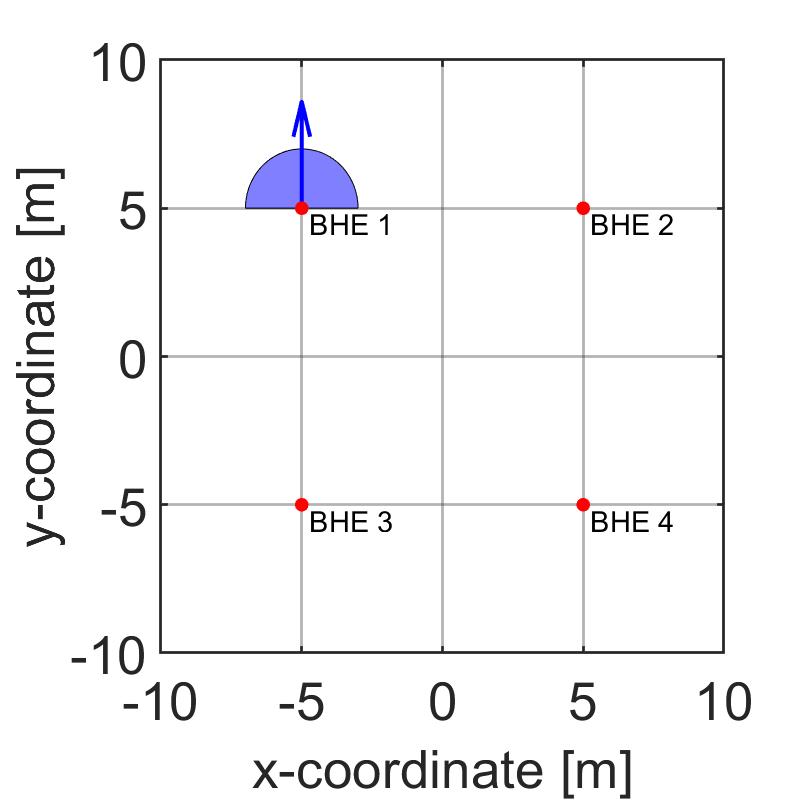}
            %\caption[Network2]%
            %{{\small Network 1}}
            %\label{fig:mean and std of net14}
        \end{subfigure}
        %\hfill
        \begin{subfigure}[b]{0.35\textwidth}
            \centering
            \includegraphics[width=\textwidth]{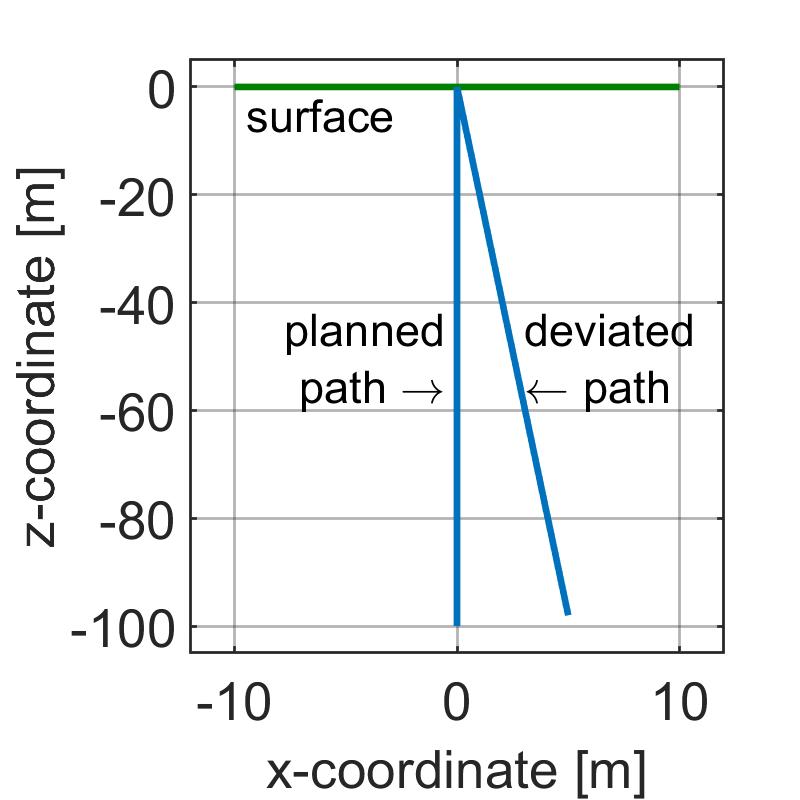}
            %\caption[]%
            %{{\small Network 2}}
            %\label{fig:mean and std of net24}
        \end{subfigure}
            \caption{Visualization of the effects of the introduced random variables. Left: A reference direction (arrow) and a range of allowed directions (blue semi-circle) of deviations. Right: Effect of a realization for the vertical angle of deviation.}
            \label{fig:dev_ex}
    \end{figure}

Let $\mathcal{X}_{\text{dir}}$ be a random variable with a probability density function (pdf) $\rho_{\text{dir}}$ on the support $[\alpha_{\text{min}},\alpha_{\text{max}}]$ governing the \textit{horizontal direction of deviation} of a BHE (azimuth). Based on a reference direction $d_{\text{ref}}$ in the $(x,y)$-plane, we allow for directions of deviations in the range of $[\alpha_{\text{min}},\alpha_{\text{max}}]$ to occur. For a realization $\alpha \in [\alpha_{\text{min}},\alpha_{\text{max}}]$, the direction of deviation is given by $d = d_{\text{ref}} +( \sin(\alpha),\cos(\alpha))^T$. A visual aide for how this is applied can be seen in Figure \ref{fig:dev_ex} on the left. Here, we see a top-down view of 4 BHEs. The first BHE may be linearly deviated in a direction as indicated by the blue semi-circle. The corresponding reference direction is $d_{\text{ref}}= (0,1)^T$ and $\text{sup}(\rho_{\text{dir}}) = [-90\degree, 90\degree]$.

Let $\mathcal{X}_{\text{dev}}$ be a random variable with a pdf $\rho_{\text{dev}}$ on the support $[0,\beta]$ governing the \textit{vertical angle of deviation} from the planned perpendicular path of a BHE (inclination). A realization of $\mathcal{X}_{\text{dev}}$ equal to zero means that the original perpendicular path is used. For any other realization, the path of the BHE will be slanted by that angle in a direction as dictated by $\mathcal{X}_{\text{dir}}$. The effect of a non-zero realization of $\mathcal{X}_{\text{dev}}$ on a BHE path is sketched in Figure \ref{fig:dev_ex} on the right.

Let $n$ be the number of considered BHEs. This gives $2n$ random variables:
%\begin{align}
%	\mathcal{X}_{i,\text{dir}}\text{ with probability density function }\rho_{i,\text{dir}}\text{ on the support }\Gamma_{i,\text{dir}} = [\alpha_{\text{i,min}},\alpha_{\text{i,max}}]~~~,~ i = 1,\dots, n  \\
%	\mathcal{X}_{i,\text{dev}}\text{ with probability density function }\rho_{i,\text{dev}}\text{ on the support }\Gamma_{i,\text{dev}} = [0,\beta_i] ~~~,~ i = 1,\dots, n
%\end{align}
\begin{align}
	\begin{aligned}
	\mathcal{X}_\text{dir}^{i}\text{ with pdf }\rho_\text{dir}^{i}\text{ on the support }\Gamma_\text{dir}^{i} = [\alpha_\text{min}^{i},\alpha_\text{max}^{i}]~~~,~ i = 1,\dots, n,  \\
	\mathcal{X}_\text{dev}^{i}\text{ with pdf }\rho_\text{dev}^{i}\text{ on the support }\Gamma_\text{dev}^{i} = [0,\beta^{i}] ~~~,~ i = 1,\dots, n.
	\end{aligned}
\end{align}

In practice, we use identical, stochastically independent random variables, so we omit the index $i$ when convenient.

So far, we have not specified the pdfs of the random variables, only their support. We are free in their choice and a concrete example is given in section \ref{uncertainty_input}.

For every joint realization $y \in \Gamma = \Gamma_\text{dir}^{1}\times~\cdots~\times~\Gamma_\text{dir}^{n} \times \Gamma_\text{dev}^{1}\times~\cdots~\times~\Gamma_\text{dev}^{n}$ of the random variables, a new geometry for the BHE array is encoded. It is possible that certain realizations produce geometries with overlapping borehole paths. In reality, these cases would constitute a critical implementation error and can not be covered by our model in a sensible manner. In order to avoid these cases, all realizations that would produce an overlap are slightly modified to the next closest viable geometry. This is done by further rotating the path of those deviated BHEs which produce an overlap until a minimum distance between paths is achieved. The minimum distance is the one recommended between BHE nodes and neighbouring nodes suggested in \cite{diersch2013feflow}, which also considers a 1D-3D coupling to model BHEs using a TRCM. Of course, this has implications for the statistical analysis, which will be discussed in section \ref{sec:discussion}.

If we want to study the performance of a particular geometry, we theoretically have to create a FEM model with it and run a simulation. However, even a single time-dependent FEM solution requires a lot of computational effort and it gives us insight into the array's performance with that particular geometry, which is only a glimpse when compared to the infinite range of geometries encoded in the random variables.

To tackle this problem, we will introduce a stochastic collocation method in the following subsection, which allows us to generate statistical information across all possible geometries while only needing to solve a few FEM problems for a selection of realizations of the random variables.

\subsection{Adaptive, anisotropic stochastic collocation}

Stochastic collocation is a sampling method which calculates an interpolant of a function depending on the support of a finite set of random variables. The interpolant is then used to estimate statistics of the considered function, such as its mean value and higher moments. In the following, we only consider the case of a real-valued function. For the example in the context of BHE arrays provided in Section \ref{sec:example}, the function is an averaged inlet-outlet-temperature depending on the domain of all possible deviating geometries, but other quantities such as a storage coefficient are reasonable, as well. Since the averaged inlet-outlet temperature over all deviations is unknown and theoretically requires the computationally expensive solution of a PDE for every possible geometry, interpolation represents an efficient method to acquire statistical information across the entire support of the random variables.

To formalize the problem and in order to remain general, let $\Gamma = \Gamma_1~\times~\Gamma_2~\times~\cdots~\times~\Gamma_D = [a_1,b_1]~\times~[a_2,b_2]~\times~\cdots~\times~[a_D,b_D]$ be the joint support of a finite set of stochastically independent random variables $\mathcal{X}_1,~\mathcal{X}_2,~\dots~\mathcal{X}_D$. Let $u(\textbf{x}),~u:\Gamma \rightarrow \R$ be a real-valued function on that space. The aim is to produce an interpolant $\textbf{I}[u](\textbf{x}),~\textbf{I}[u]:\Gamma \rightarrow \R$ of $u(\textbf{x})$. Rectangular brackets mark the interpolated function.

To this end, we require a set of interpolation nodes $y_j \in \Gamma$, which in this context are realizations of the underlying random variables. Interpolation in multiple dimensions is not as straightforward as in a single dimension and the choice of these nodes is crucial. Arbitrary sets of nodes are not guaranteed to have an interpolation function \cite{xiu2005high}. Even if it exists, poorly chosen interpolation nodes are known to produce strongly oscillating interpolants, commonly known as Runge's phenomenon \cite{fornberg2007runge}. Interpolants exhibiting this behaviour are unfit to be used for a statistical analysis of the underlying function. Additionally, the function evaluation for each node requires the solution of a PDE in our context, so we are strongly interested in using as few as possible.
These considerations lead us to using hierarchical interpolation based on Smolyak sparse grid theory \cite{novak1996high, smolyak1963quadrature}. The concepts discussed in the following sections were implemented in the software FastCOIN, which is based on the work in \cite{schieche2012unsteady}. FastCOIN was used for the stochastic collocation performed in Section \ref{sec:example}.

\subsubsection{Smolyak sparse grids}
Smolyak sparse grids use tensor products of one-dimensional interpolation rules to construct interpolation rules for multiple dimensions. In order to describe their structure, let $X := \{y_j\}_{j=1}^{m}$ be a set of interpolation nodes for one dimension. The corresponding one-dimensional interpolation scheme with Lagrangian basis polynomials for a function $u(x)$ is
\begin{align}
\label{def_1Dinterpolation}
 	I[u](x) = \sum_{j=1}^{m} u(y_j) L_j(x).
\end{align}
With this, we can define tensor product interpolation rules for a function $u(\textbf{x}) = u(x_1,\dots,x_D)$ in multiple dimensions:
\begin{align}
\label{def_MDinterpolation}
\left(I^1\otimes ~ \cdots ~ \otimes I^D\right)[u](\textbf{x}) := \sum_{j_1=1}^{m_{1}}\cdots\sum_{j_D=1}^{m_{D}} u(y_{j_1},\dots,y_{j_D})L_{j_1}(x_1)\cdots L_{j_D}(x_D).
\end{align}
These full tensor products are theoretically valid interpolation rules, however they require a number of interpolation nodes which increases exponentially with the dimension $D$. This makes them far too inefficient and we only use them here to introduce the notation.

Next, we consider sequences of one-dimensional interpolation rules $(I^{i_d}[u])_{i_d \in \N}$ for each dimension $d = 1,~\dots~D$. Definitions (\ref{def_1Dinterpolation}) and (\ref{def_MDinterpolation}) are adjusted accordingly to
\begin{align}
I^{i_d}[u](x) = \sum_{j=1}^{m_{i_d}} u(y^{i_d}_j)L^{i_d}_j(x).
\end{align}
and
\begin{align}
\left(I^{i_1}\otimes ~ \cdots ~ \otimes I^{i_D}\right)[u](\textbf{x}) := \sum_{j_1=1}^{m_{i_1}}\cdots\sum_{j_D=1}^{m_{i_D}} u(y_{j_1}^{i_1},\dots,y_{j_D}^{i_D})L_{j_1}^{i_1}(x_1)\cdots L_{j_D}^{i_D}(x_D).
\end{align}
We assume that the interpolation rules in each sequence increase in accuracy and define the difference between two successive rules as
\begin{align}
\label{def_1D_interpolation_dif}
\triangle^{i_d} [u](x) := I^{i_d}[u](x) - I^{{i_d}-1}[u](x)
\end{align}
with $I^0 = 0$. $\triangle^{i_d}$ can be interpreted as an update from one interpolation rule to the next better one. With this, Smolyak sparse grid interpolation is defined by
\begin{align}
\label{def_smolyak}
S(k,D)[u](\textbf{x}) := \sum_{|\textbf{i}|\leq k + D}^{} \left( \triangle^{i_1}\otimes~\cdots~\otimes \triangle^{i_D} \right)[u](\textbf{x}).
\end{align}
The multi-index $\textbf{i} := (i_1,\dots,i_D)$ identifies an interpolation rule for each dimension. $|\textbf{i}|:= i_1 +~ \cdots ~+ i_D$ as per usual multi-index definition. $k \in \N$ is defined as the level of the Smolyak sparse grid and starts with $0$. For a specific $k$, the union of all points $(y_{j_1}^{i_1},\dots,y_{j_D}^{i_D})$, for which $u$ is evaluated in (\ref{def_smolyak}), defines a set of interpolation nodes in $D$ dimensions. They constitute the basic Smolyak sparse grid and we call them collocation points in this context.

In order to remain general, we have described the structure of Smolyak sparse grids without referring to any specific one-dimensional interpolation rules. We are therefore free in our choice of these rules and can pick them based on the problem at hand.

Our goal is to use the interpolant in order to investigate statistical properties of the underlying function. Statistical properties such as the mean and higher moments are integrated values over the entire support of the random variables. It is therefore sensible to use interpolation rules, which also constitute good quadrature rules. For this reason, we use Clenshaw-Curtis nodes \cite{clenshaw1960method}, which are widely utilized and have been reliable in quadrature applications \cite{foo2008multi,ganapathysubramanian2007sparse}. Clenshaw-Curtis nodes are equal to the extrema of Chebyshev polynomials. For a quadrature rule for the interval $[-1, 1]$ with $m_i$ nodes, the nodes are thus given by $X^i :=\left\{-\cos\left(\frac{\pi (j-1)}{m_i -1}\right)\right\}_{j=1}^{m_i}$. They can be linearly scaled to fit any interval $[a,b]$.

\begin{figure}[H]
	%\begin{figure}[htbp]
	\centering
	\begin{subfigure}[b]{0.24\textwidth}
		\centering
		\includegraphics[width=\textwidth]{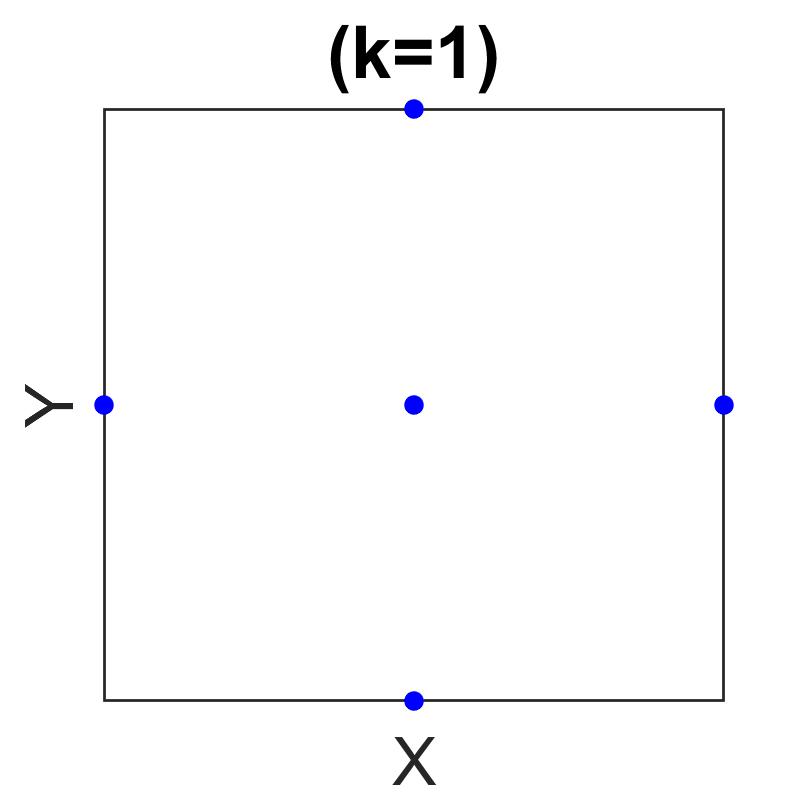}
	\end{subfigure}
	\hfill
	\begin{subfigure}[b]{0.24\textwidth}
		\centering
		\includegraphics[width=\textwidth]{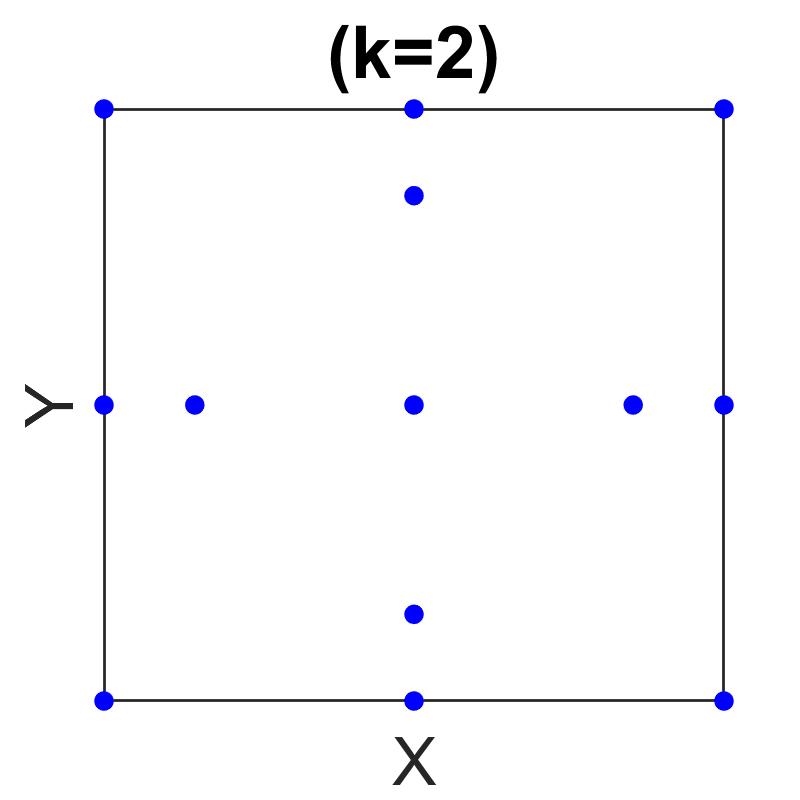}
	\end{subfigure}
	\hfill
	\begin{subfigure}[b]{0.24\textwidth}
		\centering
		\includegraphics[width=\textwidth]{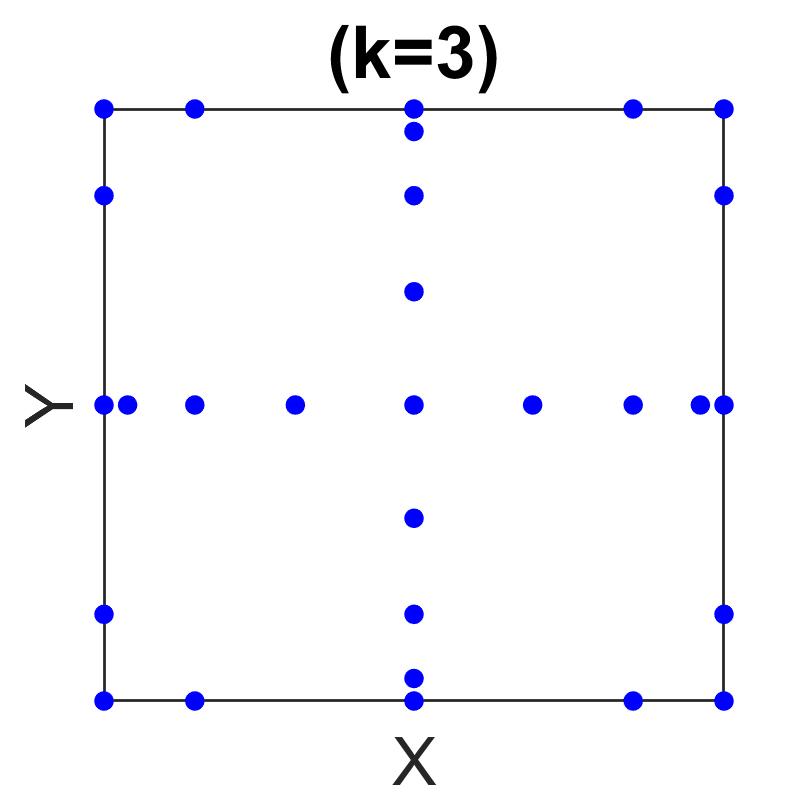}
	\end{subfigure}
	\hfill
	\begin{subfigure}[b]{0.24\textwidth}
		\centering
		\includegraphics[width=\textwidth]{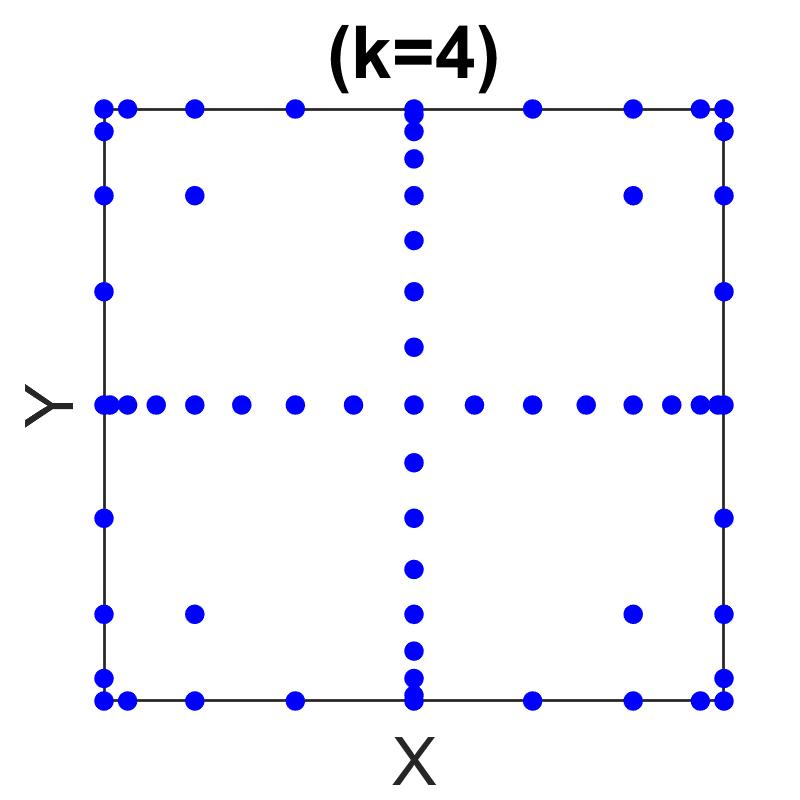}
	\end{subfigure}
	\caption{Nested Smolyak sparse grids for interpolation in two dimensions sorted by their level.}
	\label{fig:smolyak_by_level}
	% Option offen für Vergleich sparse vs full tensor
\end{figure}

Specifically, we increase the number of interpolation/quadrature nodes used in each dimension according to $m_i = 2^{i-1} + 1$. This gives a sequence of interpolation rules increasing in accuracy as demanded in (\ref{def_1D_interpolation_dif}). Secondly, this results in so called 'nested' Smolyak sparse grids, i.e. all nodes used in a grid of level $k$ are a subset of every grid of level greater than $k$. Figure \ref{fig:smolyak_by_level} illustrates this for the two-dimensional case.

With this choice, the nodes of each Smolyak sparse grid are also a subset of the nodes of the full tensor grid using the most accurate interpolation rule of the sequence in each dimension. However the total number of nodes for moderate $k$ and $D$ is much lower for the sparse grid when compared to the full tensor grid \cite{novak1999simple, schieche2012unsteady}.

The usage of nested grids leads us to hierarchical interpolation and adaptivity, which will be discussed in the following sections.

\subsubsection{Hierarchical interpolation and error estimation}
When trying to recreate an unknown function with an interpolant, we usually do not know how many nodes are needed in order to be sufficiently accurate. For this reason, we want error estimators, or at least error indicators, to judge the quality of a computed interpolant and to adaptively improve it as necessary to satisfy a given error tolerance.

The first step in achieving this is the aforementioned use of nested grids. With nested grids, we can construct an error estimator via hierarchical interpolation. Also, as we adaptively add more nodes to the interpolation rule, the calculated function evaluations at all previous nodes are reused. This is especially important in our context, as each nodal function evaluation means solving a computationally expensive PDE.

To switch to hierarchical interpolation, we make use of the nested Clenshaw-Curtis interpolation rules as specified previously, with the number of nodes increasing by $m_i = 2^{i-1} + 1$. For a sequence of one-dimensional interpolation rules according to this scheme, the corresponding sequence of interpolation nodes has the property $X^{i-1} \subseteq X^{i}$, which carries over to the Smolyak sparse grids when sorted by level.

\begin{figure}[H]
	%\begin{figure}[htbp]
	\centering
	\begin{subfigure}[b]{0.32\textwidth}
		\centering
		\includegraphics[width=\textwidth]{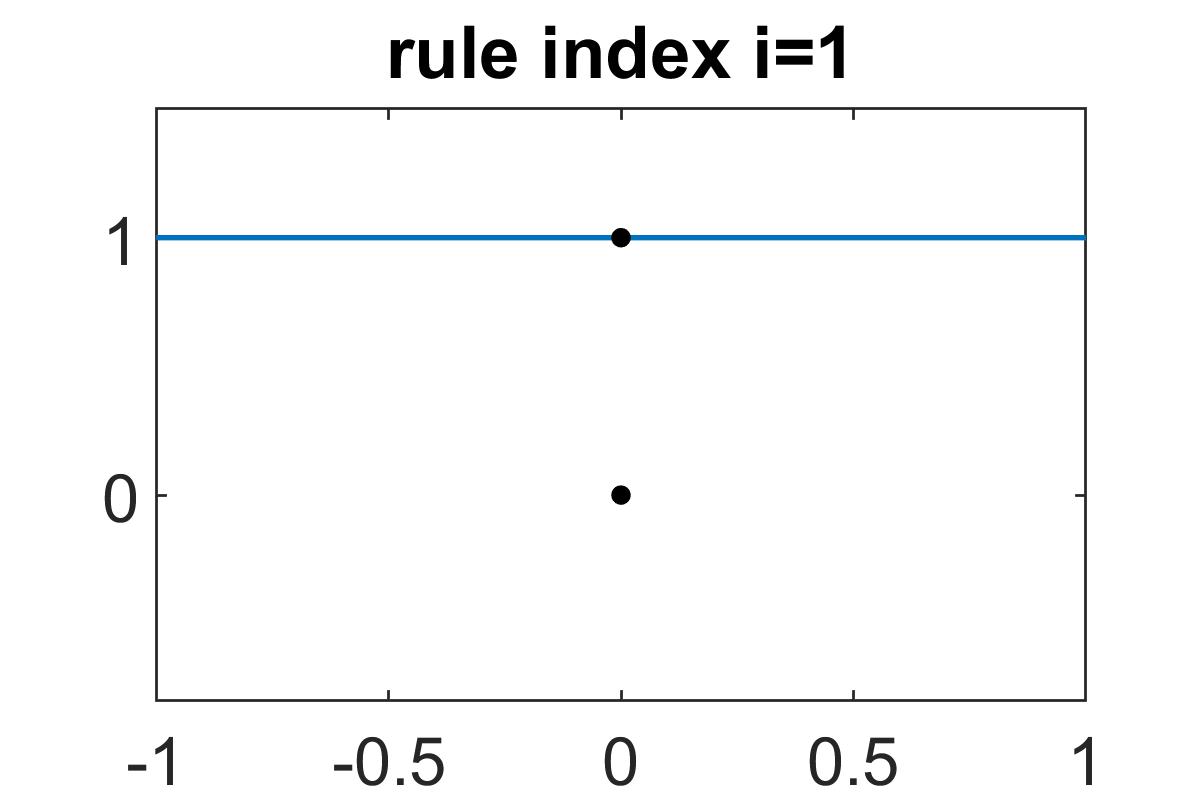}
	\end{subfigure}
	\hfill
	\begin{subfigure}[b]{0.32\textwidth}
		\centering
		\includegraphics[width=\textwidth]{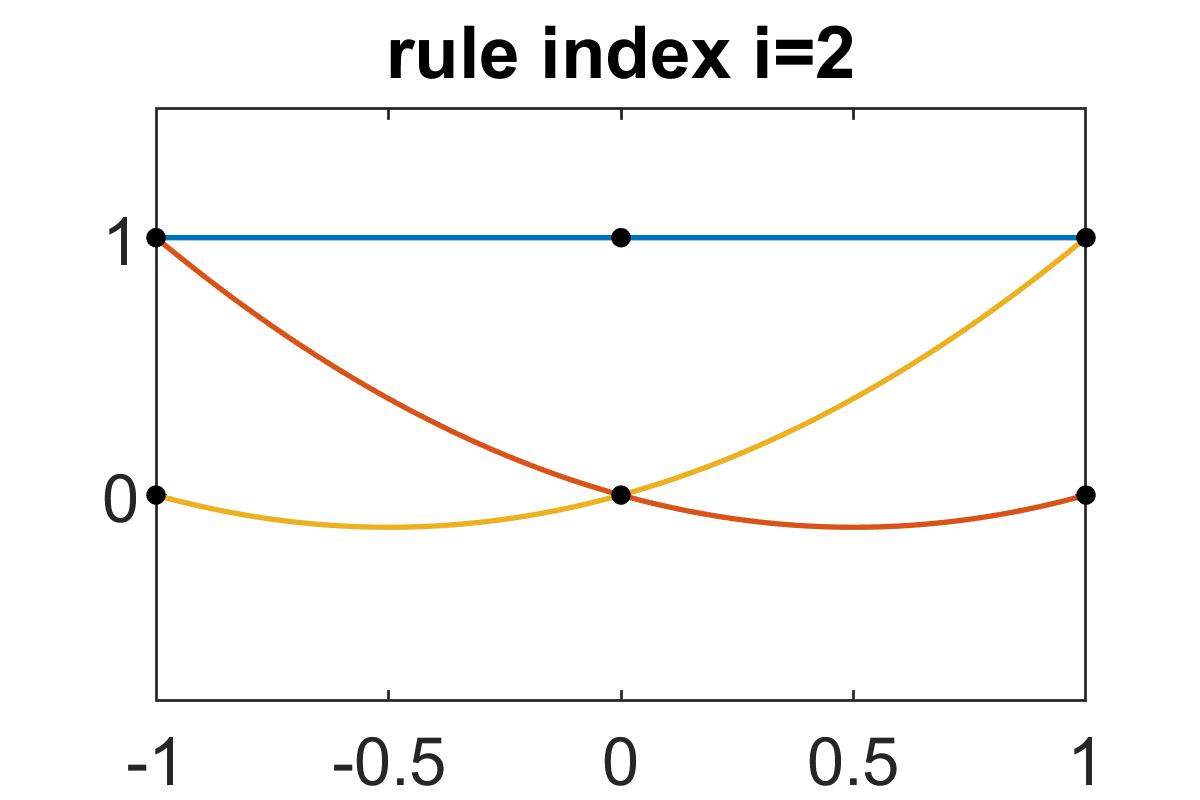}
	\end{subfigure}
	\hfill
	\begin{subfigure}[b]{0.32\textwidth}
		\centering
		\includegraphics[width=\textwidth]{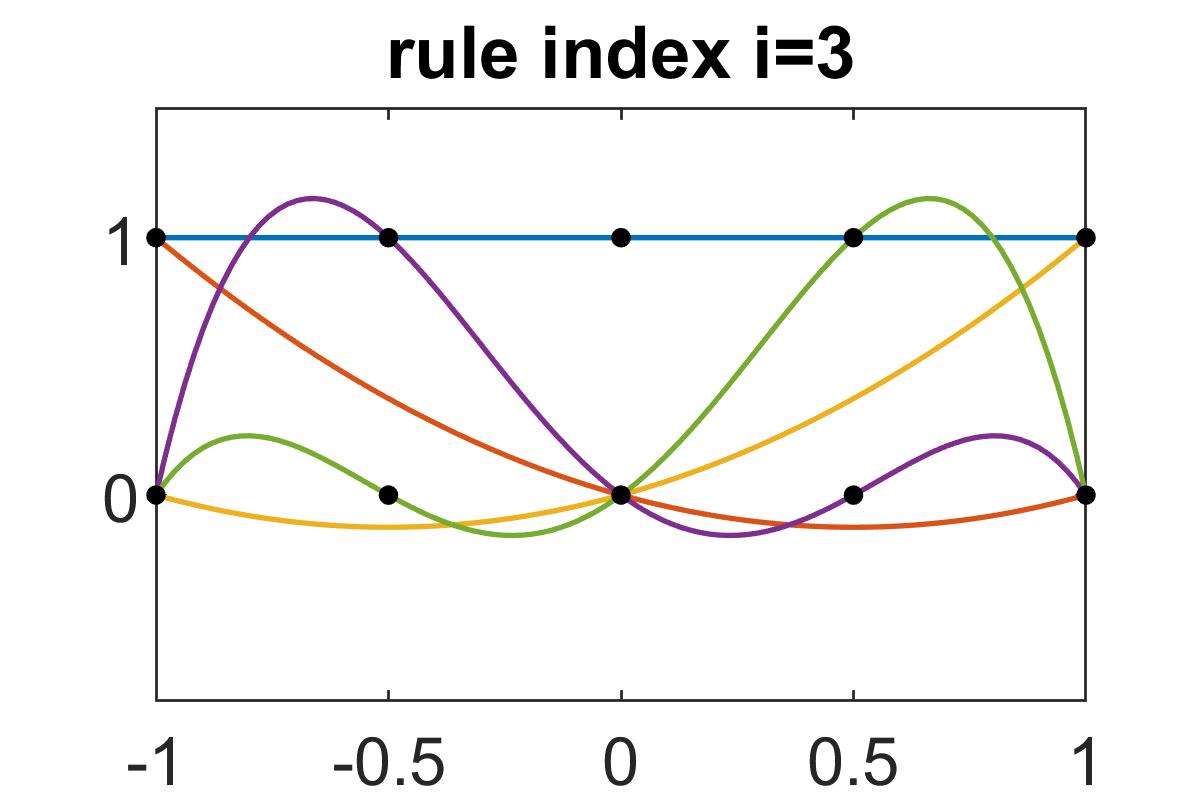}
	\end{subfigure}
	\hfill
	\caption{Hierarchical one-dimensional Lagrangian basis polynomials for a nested sequence of nodes.}
	\label{fig:hier_base_pols}
	% Option offen für Vergleich hierarchisch vs nodal
\end{figure}

We define $Y^i := X^{i}\setminus X^{i-1}$ as the nodes first appearing in $X^{i}$. For nodes $y_j^i \in Y^i$, we use Lagrangian basis polynomials that satisfy $L_{\underline{j}}^i(y_{j}^i) = \delta_{\underline{j},j}$, with $\delta_{\underline{j},j}$ being the Kronecker delta. For the remaining nodes $y_j^i \notin Y^i$, we utilize the base polynomials of the previous set, $L_j^i(x) = L_j^{i-1}(x)$. This iterative definition gives a hierarchical structure in which basis polynomials for a set $X^{i}$ do not need nodal information of sets of a higher index. The resulting basis polynomials are illustrated in Figure \ref{fig:hier_base_pols}.

With this, we can iteratively redefine a sequence of one-dimensional interpolation rules by
\begin{align}
\tilde{I}^i[u]:=\tilde{I}^{i-1}[u] + \sum_{y_j^i \in Y^i} \underbrace{\left((u - \tilde{I}^{i-1}[u])(y_j^i)  \right)}_{=:HS_j^i}L_j^i(x)
\end{align}
with $\tilde{I}^1[u] := I^1[u]$. Note that this is effectively just a change in notation and we get $\tilde{I}^i[u] = I^i[u]$ for all $i$. With this notation, we can define the hierarchical surpluses $HS_j^i$, which measure the contribution of each new node in $\tilde{I}^i[u]$ that was not in $\tilde{I}^{i-1}[u]$. In other words, they are indicators of how much each new node improves the interpolant.

Analogously, Smolyak's sparse grid interpolation formula (\ref{def_smolyak}) can be rewritten as
\begin{align}
\label{smolyak_HS}
\displaystyle
S(k,D)[u](\textbf{x}) = S(k - 1,D)[u](\textbf{x}) + \sum_{
	\substack{\textbf{y}_{\textbf{j}}^{\textbf{i}} \in Y^{i_1}~\otimes~\cdots~\otimes Y^{i_D}\\
		|\textbf{i}|= k + D}
} \underbrace{\left((u - S(k - 1,D)[u])(\textbf{y}_{\textbf{j}}^{\textbf{i}})  \right)}_{=:\textbf{HS}_{\textbf{j}}^{\textbf{i}}}\textbf{L}_{\textbf{j}}^{\textbf{i}}(\textbf{x})
\end{align}
with $\textbf{y}_{\textbf{j}}^{\textbf{i}} = (y_{j_1}^{i_1},\dots,y_{j_D}^{i_D})$ and $\textbf{L}_{\textbf{j}}^{\textbf{i}} = L_{j_1}^{i_1}\cdots L_{j_D}^{i_D}$. The hierarchical surpluses $\textbf{HS}_{\textbf{j}}^{\textbf{i}}$ serve a similar purpose as in the one-dimensional case. It is reasonable to assume, that if the hierarchical surplus of a new set of nodes in a certain dimension is large, we should invest further nodes in that dimension until the new surpluses produce only minor changes to the interpolant. We therefore use them in this way as error indicators for the Smolyak sparse grid interpolation.

\subsubsection{Adaptivity}

With the introduction of the hierarchical surpluses as error indicators, we can formulate an adaptive version of the Smolyak sparse grid interpolation. For the most basic version, we could simply increase the Smolyak-level $k$ iteratively as long as at least one hierarchical surplus is above a chosen threshold. This leads to a refinement in every dimension simultaneously and improves the interpolant uniformly. However, in practice this is inefficient, as the number of nodes required increases very quickly even for the Smolyak sparse grid when considering higher dimensions $D$. Also, it ignores the possibility that only certain dimensions require a refinement, leading to wasted effort on improving the interpolant for variables, in which it is already sufficiently accurate.

A major feature of the hierarchical surpluses is the fact that they provide an error indicator for every individual interpolation rule introduced to the Smolyak formula. Using this information, it is much more efficient to improve on the specific interpolation rules where the error indicator is largest than to refine the grid uniformly.

To this end, we initiate the interpolation with the most basic Smolyak sparse grid corresponding to the smallest multi-index $\textbf{i} := (1,\dots,1)$. Starting with this index, we iteratively add more indices, giving an index-set $F$, and their corresponding interpolation rules to the grid. As mentioned earlier, each index relates to an interpolation rule with its accompanying hierarchical surplus.

The process of adding more indices is done as follows: We select the multi-index $\textbf{i}_R$ of $F$ with the highest hierarchical surplus to be refined. To this index, we separately add $1$ to each dimension, giving us $D$ new potential indices. Each of these indices first has to be checked for admissibility. If a new potential index $\textbf{i} := (i_1,\dots,i_D)$ satisfies $\textbf{i} - e_d \in F$ ($e_d $ being the unit vector of dimension $d$) for all $d=1,\dots,D$ with $i_d > 1$, we add it to the index set and include its related interpolation rule to the formula. The admissibility criterion makes sure that the hierarchical structure of the grid is preserved. An added index and its interpolation rule in turn give a new hierarchical surplus, which again might indicate that more refinement to that index is necessary. The refined index $\textbf{i}_R$ and its hierarchical surplus become inactive. All unrefined indices are considered active by default. This scheme is performed iteratively until all active indices with a hierarchical surplus greater than a chosen threshold have been refined and none of the active hierarchical surpluses violate the threshold. Alternatively, the sum of all active hierarchical surpluses can serve as an error indicator, which more naturally judges the overall quality of the interpolant.

The resulting dimension-adaptive, anisotropic Smolyak algorithm is called the generalised Smolyak algorithm and was introduced in \cite{gerstner2003dimension}.

\subsubsection{Statistics}
\label{sec:Statistics}

The calculated interpolant $\textbf{I}[u](\textbf{x})$ allows us to estimate various statistical quantities of the original function $u(\textbf{x})$, if it is sufficiently accurate. This is ensured through the adaptive process using error indicators.

Using a simplified notation of (\ref{smolyak_HS}), we write
\begin{align}
u(\textbf{x}) \approx \textbf{I}[u](\textbf{x}) = \sum_{j = 1}^p \textbf{HS}_{\textbf{j}}\textbf{L}_{\textbf{j}}(\textbf{x}).
\end{align}
Instead of summating as indicated by the sparse grid structure, we sum over all collocation points used in the formula. This leads to a very simple expression for the interpolant after having calculated the hierarchical surpluses. We define $p$ as the total number of collocation points and accordingly rearrange $\{ \textbf{y}_{\textbf{j}} \}_{j=1}^{p}$, $\{ \textbf{HS}_{\textbf{j}} \}_{j=1}^{p}$ and $\{ \textbf{L}_{\textbf{j}} \}_{j=1}^{p}$.

Using this expression, we can straightforwardly calculate an estimate of the mean of $u(\textbf{x})$. Let therefore $\rho(\textbf{x}) = \prod_{i=1}^D  \rho_i(x_i)$ be the joint density of $\mathcal{X}_1,~\mathcal{X}_2,~\dots~\mathcal{X}_D$. Then
\begin{align}
E[u(\textbf{x})] \approx \int_{\R^D} \sum_{j = 1}^p \textbf{HS}_{\textbf{j}}\textbf{L}_{\textbf{j}}(\textbf{x})\rho(\textbf{x}) ~d\textbf{x}= \sum_{j = 1}^p \textbf{HS}_{\textbf{j}}E\left[ \textbf{L}_{\textbf{j}}(\textbf{x})\right].
\end{align}
The calculation of $E\left[ \textbf{L}_{\textbf{j}}(\textbf{x})\right]$ is computationally inexpensive for most common pdfs $\rho$. Similarly, we can estimate higher moments of $u(\textbf{x})$ as well.

Instead of estimating the mean E[u(\textbf{x})], which is an integration over the entire support, we can also choose to only partially integrate in order to investigate the effect of specific random variables.

Lastly, through the use of Monte Carlo sampling of the interpolant and kernel density estimation, we can estimate the pdf of $u$ itself.

An application of these techniques will be presented  in section \ref{sec:example}.

\section{Case study}
\label{sec:example}

In this section, we apply the previously discussed model and methods in a case study in order to investigate the performance of a BHE array in a heat extraction scenario. The original array layout, soil properties and operation regime were designed using Earth Energy Designer (EED) \cite{hellstrom2000earth} and constitute a realistic application. The BHE array is considered in three base geometries, each of which is investigated in 6 subsequent uncertainty quantification scenarios with randomly deviating borehole paths. We showcase the statistical information that the method yields and discuss the results.

\subsection{Uncertainty quantification setup}

\subsubsection{FEM model}
\label{FEM_model}

We consider a cuboid patch of soil bordering the earth's surface at the top with a horizontal extent of 120m x 120m and a depth of 120m. At the top and the bottom, the temperature is fixed via a Dirichlet boundary condition of $10~\celsius$ and $13.6~\celsius$ respectively. Throughout the cuboid, the initial temperature is set as the equilibrium dictated by both boundary conditions. This conforms to the natural geothermal gradient in an undisturbed patch of soil. At every other boundary, a zero Neumann boundary condition is used.

\begin{figure}[H]
%\begin{figure}[htbp]
	\centering
	\begin{subfigure}[b]{0.365\textwidth}
		\centering
		\includegraphics[width=\textwidth]{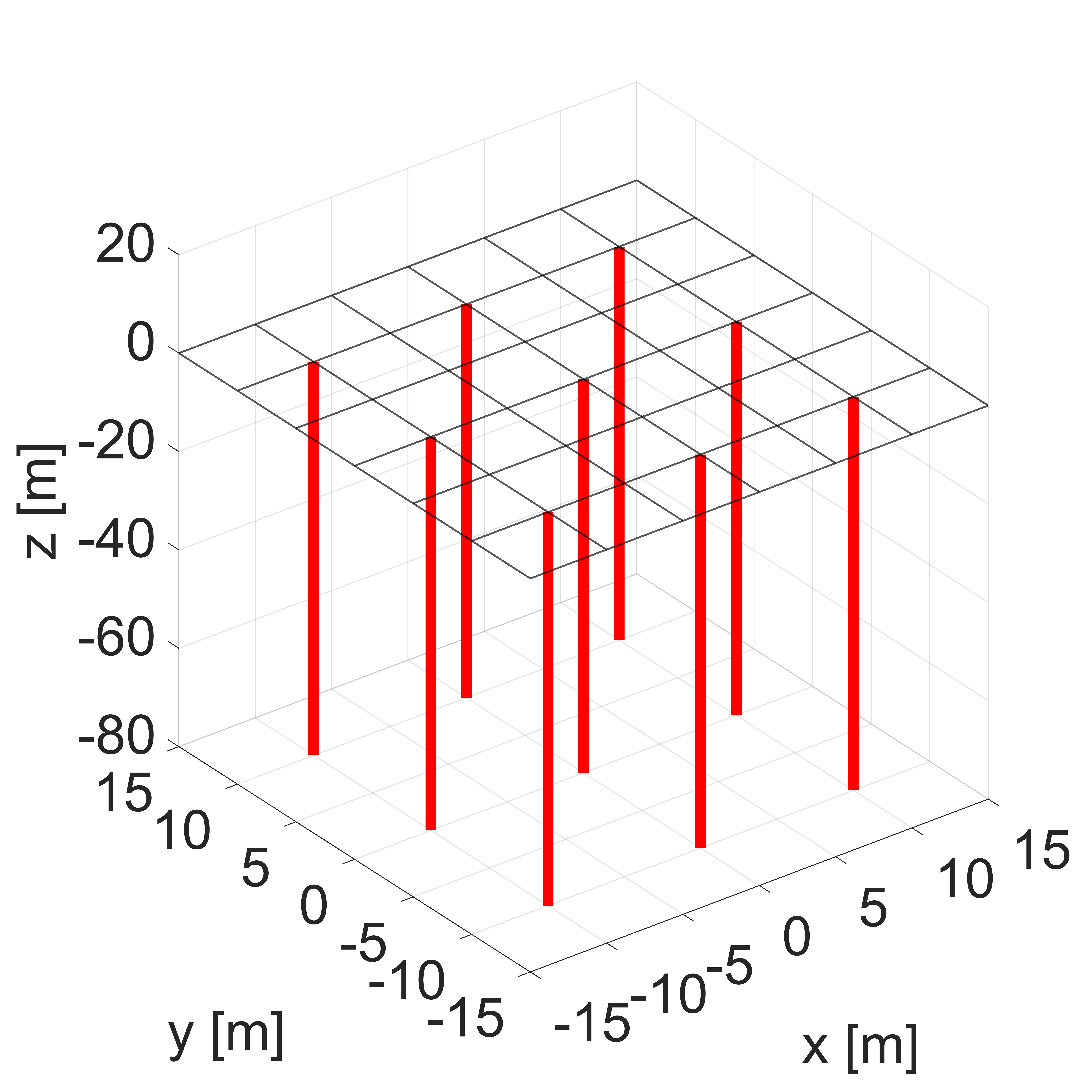}
		%\caption{Layout and placement of the BHEs in the BTES.}
		%{{\small Network 1}}
		%\label{fig:mean and std of net14}
	\end{subfigure}
	%\hfill
	\begin{subfigure}[b]{0.31\textwidth}
		\centering
		\includegraphics[width=\textwidth]{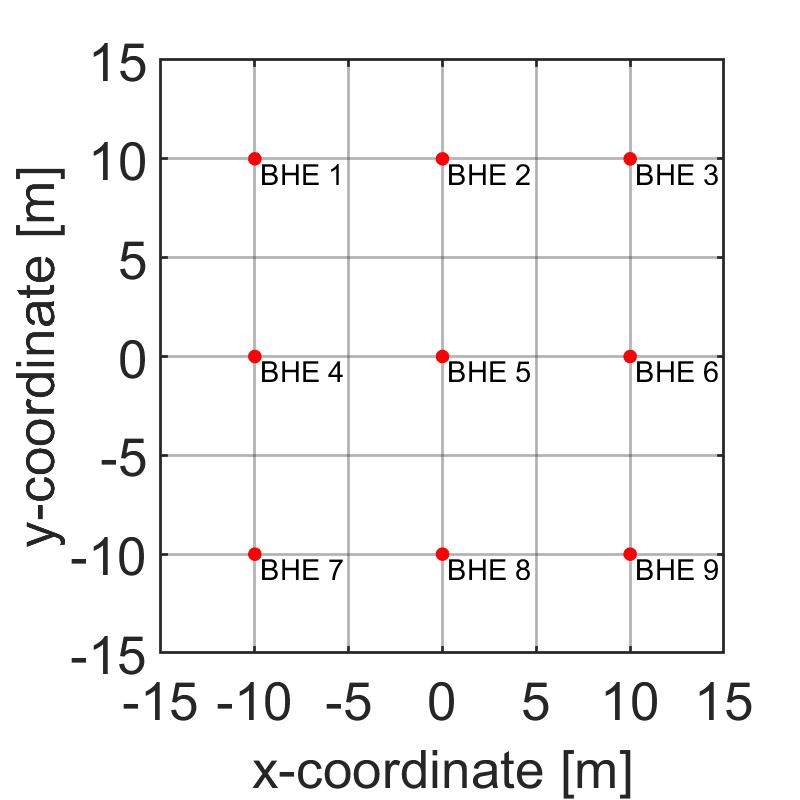}
		%\caption{Layout and placement of the BHEs in the BTES.}
		%{{\small Network 1}}
		%\label{fig:mean and std of net14}
	\end{subfigure}
	\begin{subfigure}[b]{0.31\textwidth}
		\centering
		\includegraphics[width=\textwidth]{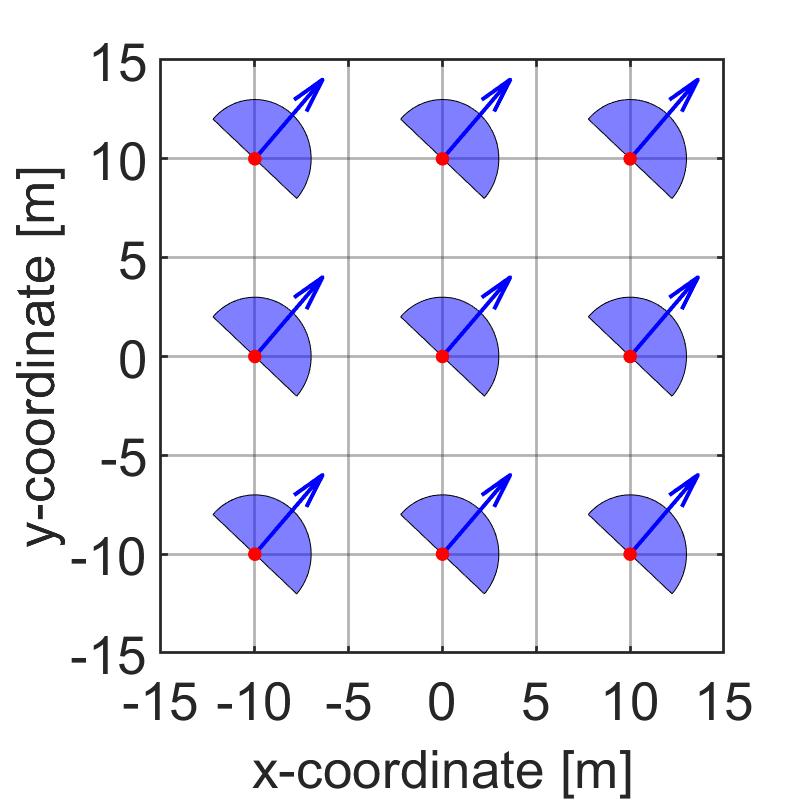}
		%\caption{Layout and placement of the BHEs in the BTES.}
		%{{\small Network 2}}
		%\label{fig:mean and std of net24}
	\end{subfigure}
	\caption{Layout, placement and directions of deviation of the BHEs in the array.}
	\label{BTES_layout}
\end{figure}

In the horizontal center, an array with 9 BHEs is simulated. We consider three base layouts for this facility. In the first, the BHEs are arranged within a 20m x 20m square. The BHEs each form a line perpendicular to the surface to a depth of 81m. For the second layout, the BHEs are placed closer within a 12m x 12m square and in the third, we place them farther apart within a 28m x 28m square. In the following, we identify each base layout by its horizontal extent: 12m, 20m and 28m. The 20m-layout was taken from the simulation software EED and is the suggested optimum when operating the BHE array as detailed in section \ref{BTES_operation}. The 12m-and 28m-layout are variations of this case in order to study the sensitivity to deviating boreholes with regards to differing base geometries. The 20m-layout is sketched in Figure \ref{BTES_layout}. The used soil parameters of the underlying system of partial differential equations are notated in Table \ref{parameters_FEM}.

\begin{table}[t!]
	%\begin{table}[htbp]
	\centering
	\begin{tabu}{|l|l|l|l|l|l|}
		\hline
		\multicolumn{4}{|c|}{model parameters} \\
		\hline
		symbol & parameter name & value & unit \\
		\hline
		$\alpha$ & temperature coefficient, fluid & $-3.0 \cdot 10^{-4}$ & $K^{-1}$ \\
		$\beta$ & compressibility, fluid & $4.45 \cdot 10^{-10}$ & $m ~ s^2 ~ (kg)^{-1}$ \\
		$c$ & specific heat capacity, fluid & $4200.0$ & $ J ~ (kg)^{-1} ~ K^{-1}$  \\
		%$\rho$ & density, fluid & $1000.0$ & $ (kg)~ m^{-3}$  \\
		$\rho$ & density, fluid & $977.0$ & $ (kg)~ m^{-3}$  \\ %
		$\mu$ & viscosity, fluid & $10^{-3}$ & $(kg)~ m^{-1} ~ s^{-1}$ \\
		$ n $ & porosity & $ 0.2 $ &  \\
		$ k $ & permeability & $ 1.019 \cdot 10^{-15}$ & $m^2$ \\
		$\lambda_L$ & longitudinal heat conductivity  & $2.1 \cdot 10^{7}$ & $ J ~ m^{-2} ~ K^{-1}$ \\
		$\lambda_T$ & transverse heat conductivity  & $2.1 \cdot 10^{6}$ & $ J ~ m^{-2} ~ K^{-1}$  \\
		$\rho^s c^s$ & vol. heat capacity, solid/rock & $ 2.26 \cdot 10^6$ & $J ~  m^{-3} ~ K^{-1}$ \\
		$\kappa$ & thermal conductivity & $2.21$ & $W ~ m^{-1} ~ K^{-1}$ \\
		\hline
	\end{tabu}
	\caption{Material parameters for the FEM model.}
	\label{parameters_FEM}
\end{table}

\subsubsection{BHE array operation}
\label{BTES_operation}

\begin{figure}[H]
	%\begin{figure}[htbp]
	\centering
	%\begin{subfigure}[b]{0.42\textwidth}
	\begin{subfigure}[b]{0.535\textwidth}
		\centering
		\includegraphics[width=\textwidth]{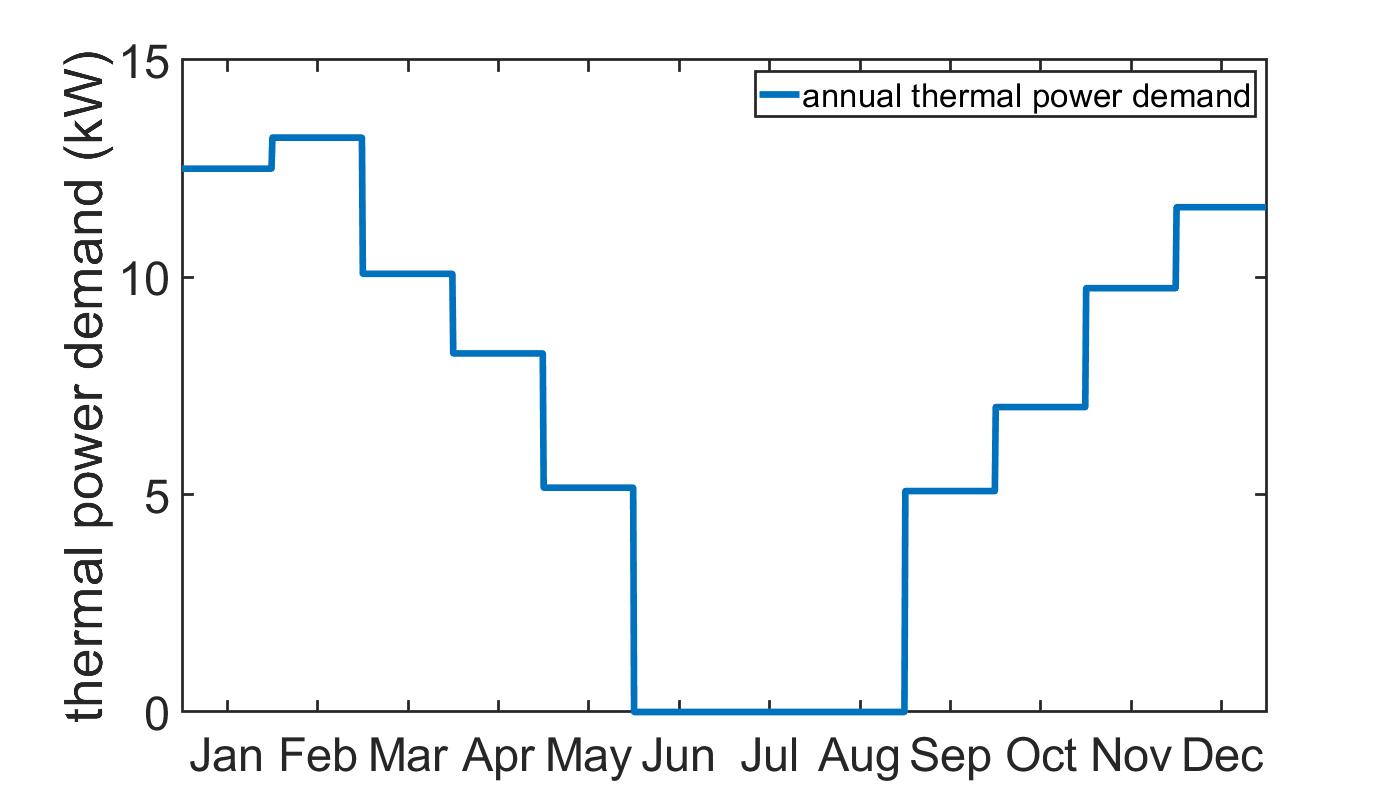}
		%\caption{Layout and placement of the BHEs in the BTES.}
		%{{\small Network 1}}
		%\label{fig:mean and std of net14}
	\end{subfigure}
	%\hfill
	%\begin{subfigure}[b]{0.36\textwidth}
	\begin{subfigure}[b]{0.455\textwidth}
		\centering
		\includegraphics[width=\textwidth]{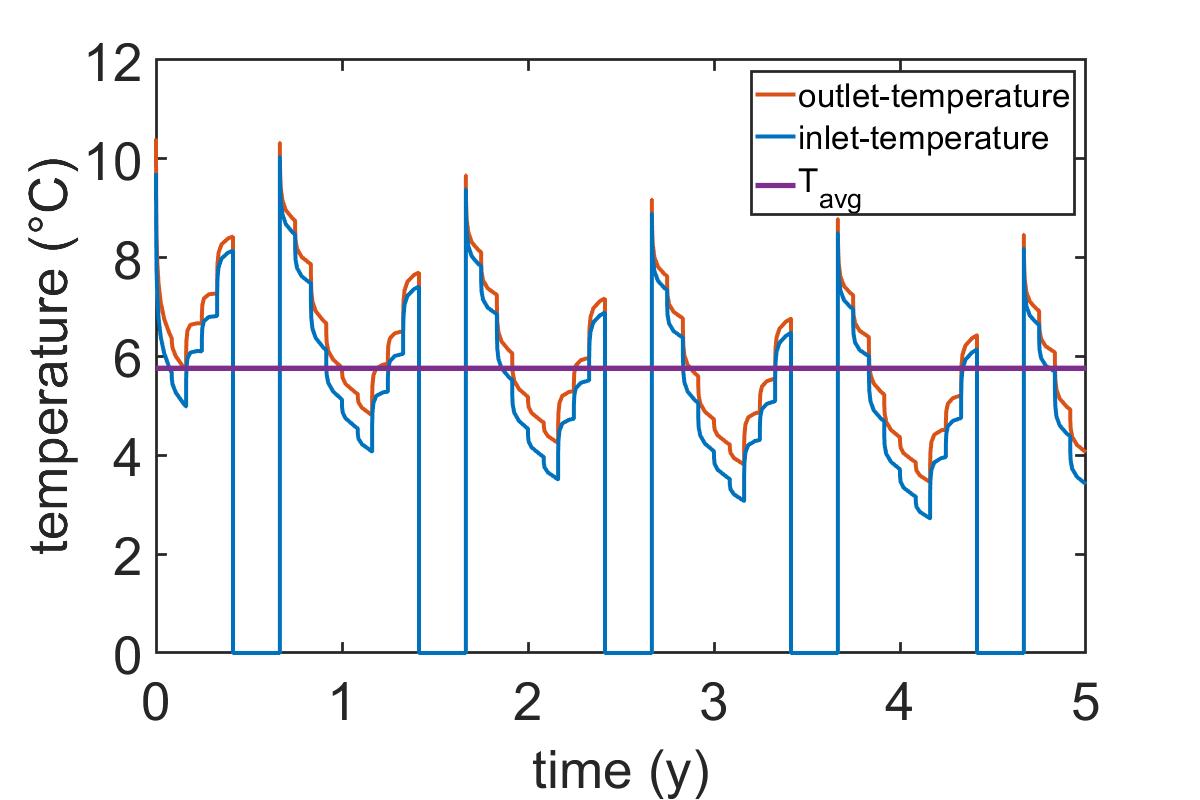}
		%\caption{Layout and placement of the BHEs in the BTES.}
		%{{\small Network 1}}
		%\label{fig:mean and std of net14}
	\end{subfigure}
	\caption{Annual thermal power demand which the BHE array has to supply (left). Resulting inlet-outlet-temperature profile for the undeviated array using the 20m-layout (right).}
	\label{BTES_annual_workload}
\end{figure}

The BHE array is operated in an extraction application for a total of 5 years with a monthly changing workload, where each month, a constant power load is extracted from the soil. The inlet for all BHEs is supplied with a fluid with identical inlet temperature and flowrate. The outlet temperature of all BHEs is averaged. The difference between the inlet and averaged outlet temperature provides the specified powerload. The inlet temperature is constantly adjusted to ensure that the current powerload target is achieved at all times. The annual workload can be seen in Figure \ref{BTES_annual_workload}. This regime represents a typical workload for a BHE array of this type and size when supplying a larger building or when connected to a heat pump in a district heating application.

All BHEs are identical and are fed the inlet fluid with a constant flowrate of 0.5 liters per second. The BHE material parameters and specifications are notated in Table \ref{parameters_BHE}.
\begin{table}[t!]
	%\begin{table}[htbp]
	\centering
	\begin{tabu}{ |l|l|l|}
		\hline
		\multicolumn{3}{|c|}{BHE material parameters} \\
		\hline
		parameter name & value & unit \\
		\hline
		BHE type & double U & \\
		length & 81 & $m$ \\
		borehole diameter & 0.1522 & $m$ \\
		outer diameter, pipe & 0.032 & $m$ \\
		wall thickness, pipe & 0.0029 & $m$  \\
		shank space of two parallel shanks & 0.072 &  $m$  \\
		thermal conductivity, pipe & 0.42 & $W~ m^{-1} ~ K^{-1} $ \\
		specific heat capacity, fluid & 3795 & $J~ (kg)^{-1} ~ K^{-1} $  \\
		thermal conductivity, fluid & 0.48 & $W~ m^{-1} ~ K^{-1} $ \\
		refrigerant dynamic viscosity, fluid & 0.0052& $(kg) ~m^{-1} ~ s^{-1} $ \\
		density, fluid & 1052 & $(kg) ~ m^{-3}$  \\
		thermal conductivity, grout  & 2 & $W~ m^{-1} ~  K^{-1} $ \\
		\hline
	\end{tabu}
	\caption{BHE material parameters and measurements.}
	\label{parameters_BHE}
\end{table}

\subsubsection{Uncertainty input and quantity of interest}
\label{uncertainty_input}

The quantity of interest that we want to investigate is the averaged inlet- and outlet-temperature $T_{avg}$ of the BHE array for the 5 years of operation, except for June, July and August, where the facility is not in use. Informally, we define
%\begin{align}
%T_{avg} := \displaystyle \frac{1}{9} \sum_{i=1}^{9}\left( \frac{\displaystyle \int_{t_{SIM}} \frac{T_{\text{in}}^i(t) + T_{\text{out}}^i(t)}{2} dt}{t_{SIM}}\right) ~~~,~ t_{SIM}:=\text{5 years} \setminus \{\text{June, July, August}\}.
%\end{align}
%\begin{align}
\begin{equation}
	\begin{aligned}
		T_{avg} :&= \displaystyle \frac{1}{9} \sum_{i=1}^{9} \frac{1}{t_{SIM}} \displaystyle \int_{t_{SIM}} \frac{T_{\text{in}}^i(t) + T_{\text{out}}^i(t)}{2} dt \\
		t_{SIM}:&=\text{5 years} \setminus \{\text{June, July, August}\}.
	\end{aligned}
\end{equation}
%\end{align}
$T_{avg}$ is in direct relation to the coefficient of performance (COP) of the BHE array, which is a measure of its effectiveness when operated with a heat pump. A high $T_{avg}$ means that the heat pump requires less electricity to supply a specified amount of heat, resulting in a higher and thus better COP.
\begin{align}
COP(T_{avg}) = \frac{T_{\text{ref}}}{T_{\text{ref}} - T_{avg}} ~~~,~\text{all temperatures in Kelvin}.
\end{align}
$T_{\text{ref}}$ is the target temperature of the heated water supplied by the heat pump. We consider $T_{\text{ref}} = 35~\celsius$.

To investigate the performance, we will be using the discussed stochastic collocation method to calculate an interpolant of $T_{avg}$ in 6 successive scenarios for each layout, in which we introduce increasingly strong sources of uncertainty into the deterministic setting described in sections \ref{FEM_model} and \ref{BTES_operation}.

In all scenarios, we assume that every individual BHE can be linearly deviated independently in the directions indicated by the semi-circles in Figure \ref{BTES_layout} on the right for the 20m-layout. The same applies to the 12m- and 28m-layout. This gives 18 variables governing the deviations, 2 for each BHE. As discussed in section \ref{modeling_uncertainty}, for each BHE, we introduce a random variable governing the direction of deviation and one governing the vertical angle of the deviation.

\begin{figure}[H]
	%\begin{figure}[htbp]
	\centering
	\begin{subfigure}[b]{0.28\textwidth}
		\centering
		\includegraphics[width=\textwidth]{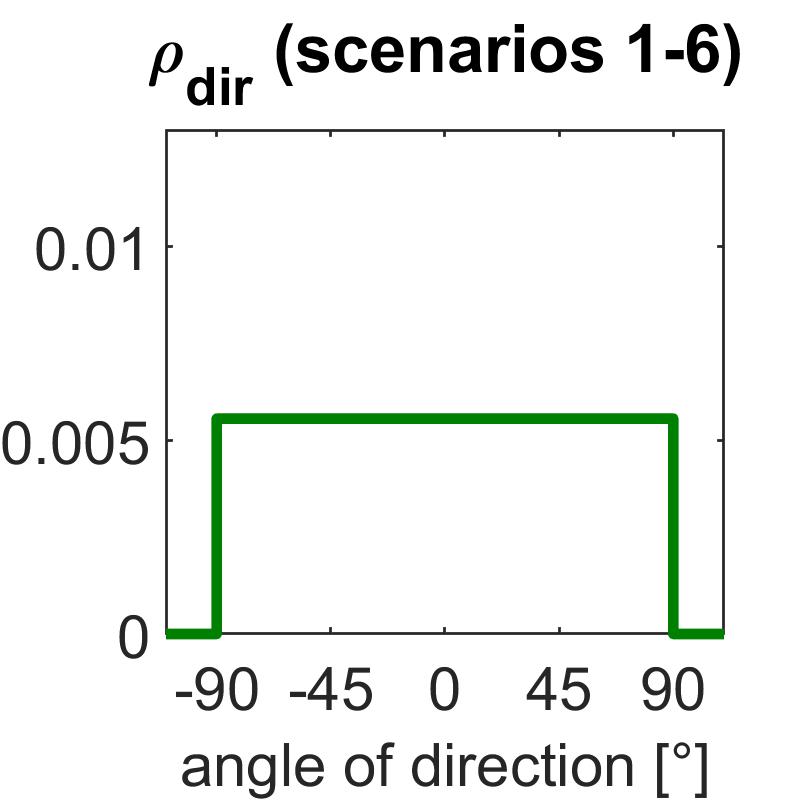}
		%\caption[Network2]%
		%{{\small Network 1}}
		%\label{fig:mean and std of net14}
	\end{subfigure}
	\hspace{4mm}
	\begin{subfigure}[b]{0.28\textwidth}
		\centering
		\includegraphics[width=\textwidth]{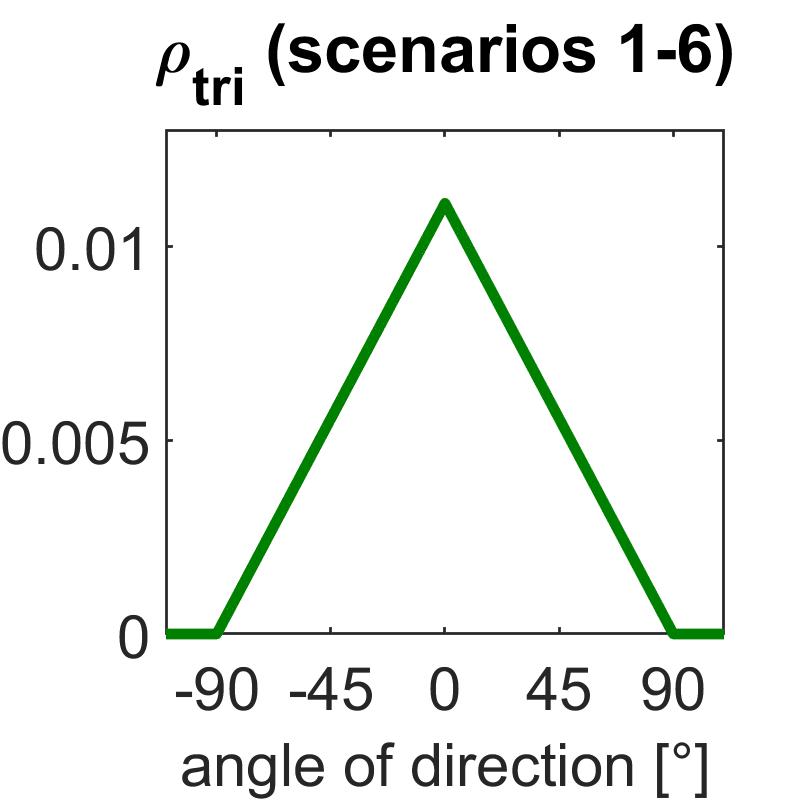}
		%\caption[Network2]%
		%{{\small Network 1}}
		%\label{fig:mean and std of net14}
	\end{subfigure}
	\vskip\baselineskip
	\centering
	\begin{subfigure}[b]{0.28\textwidth}
		\centering
		\includegraphics[width=\textwidth]{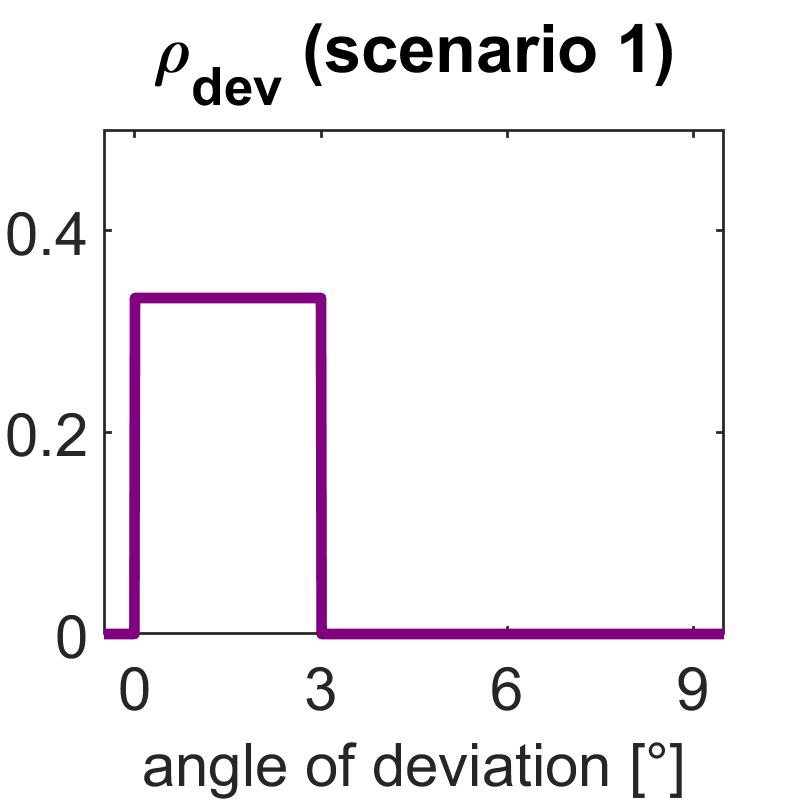}
		%\caption[]%
		%{{\small Network 2}}
		%\label{fig:mean and std of net24}
	\end{subfigure}
	\hspace{4mm}
	\begin{subfigure}[b]{0.28\textwidth}
		\centering
		\includegraphics[width=\textwidth]{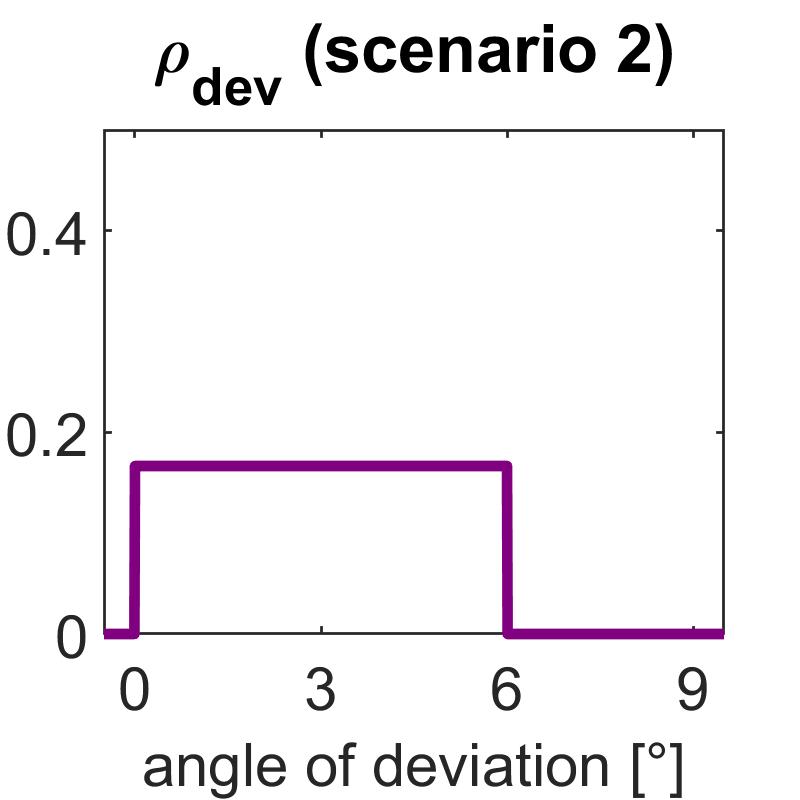}
		%\caption[]%
		%{{\small Network 2}}
		%\label{fig:mean and std of net24}
	\end{subfigure}
	\hspace{4mm}
	\begin{subfigure}[b]{0.28\textwidth}
		\centering
		\includegraphics[width=\textwidth]{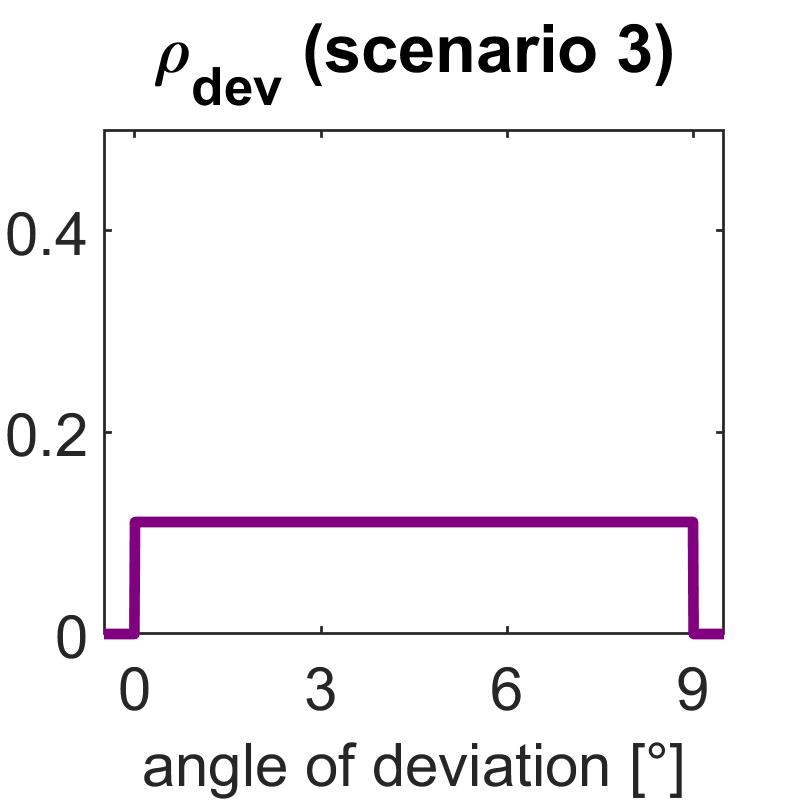}
		%\caption[]%
		%{{\small Network 2}}
		%\label{fig:mean and std of net24}
	\end{subfigure}
	%\caption{Probability density functions governing the input uncertainty.}
	\caption{Selection of pdfs used in the successive scenarios.}
	\label{fig:pdf-input}
\end{figure}

For the directions, we consider 9 identical, uniformly distributed random variables $\left \{ {\mathcal{X}}_\text{dir}^{i} \right \}_{i=1}^{9}$ with a pdf $\rho_{\text{dir}}$ on the support $\Gamma_\text{dir} $ as shown in Figure \ref{fig:pdf-input} at the top left. This corresponds to a range of possible directions of $180\degree$ with the reference direction $d_{\text{ref}} = (\cos(40\degree),\sin(40\degree))^T$ for every BHE, every direction being equally likely. $d_{\text{ref}}$ represents a shift of the direction $d = (1,1)^T = (\cos(45\degree),\sin(45\degree))\cdot \sqrt{2}$ by $-5\degree$ and is used to avoid exact overlapping of the borehole paths during sampling. Should overlapping occur for a sampled geometry, it is slightly modified to a viable one. The random variables governing the directions are kept the same in all 6 scenarios.

For the vertical angles of deviation of every BHE, we also assume 9 identical, uniformly distributed random variables $\left \{ {\mathcal{X}}_\text{dev}^{i} \right \}_{i=1}^{9}$ with a pdf $\rho_{\text{dev}}$ on the support $\Gamma_\text{dev} $. In contrast to the directions, a different set will be used in each scenario. For the first scenario, we allow for a vertical angle of up to $3\degree$. In each subsequent scenario, we increase the maximum vertical angle by another $3\degree$, up to a total of $18\degree$ in the last scenario. The corresponding pdfs $\rho_{\text{dev}}$ for the scenario 1, 2 and 3 respectively are shown in Figure \ref{fig:pdf-input}. The remaining pdfs follow the same pattern. Practically speaking, this means that with each subsequent scenario, we allow for stronger possible deviations to occur.
The complete information regarding the input uncertainty used in each scenario is summed up in Table \ref{table:scenario_pdfs}.
\begin{table}[t!]
%\begin{table}[htbp]
	\centering
	\begin{tabu}{ |l|l|l|}
		\hline
		rand. var. & support & probability density function\\
		\hline\xrowht{28pt}
		$\left \{ {\mathcal{X}}_\text{dir}^{i} \right \}_{i=1}^{9}$ &
		$\left \{\Gamma_\text{dir}^{i}  = [-90\degree, 90\degree]\right \}_{i=1}^{9}$ &
		$\left \{ \rho_{\text{dir}}^{i} (x)=\left\{
		\begin{array}{ll} 0.0056, &
		x\in \Gamma_\text{dir}^{i} \\
		0, & x\not\in \Gamma_\text{dir}^{i} \end{array}\right. \right \}_{i=1}^{9}$\\
		\hline\xrowht{28pt}
		$\left \{ {\mathcal{X}}_\text{dev}^{i} \right \}_{i=1}^{9}$
		& $\left \{\Gamma_\text{dev}^{i}  = [0\degree, (3j)\degree]\right \}_{i=1}^{9}$ &
		$ \left \{ \rho_{\text{dev}}^{i} (x)=\left\{
		\begin{array}{ll} 1/(3j), &
		x\in \Gamma_\text{dev}^{i} \\
		0, & x\not\in \Gamma_\text{dev}^{i} \end{array}\right. \right \}_{i=1}^{9}$\\
		\hline\xrowht{38pt}
		$\left \{ {\mathcal{X}}_\text{tri}^{i} \right \}_{i=1}^{9}$ &
		$\left \{\Gamma_\text{tri}^{i}  = [-90\degree, 90\degree]\right \}_{i=1}^{9}$ &
		$ \left \{ \rho_{\text{tri}}^{i} (x)=\left\{
		\begin{array}{lll}
		\frac{1}{90} + \frac{x}{90^2}, & x\in [-90\degree, 0\degree] \\
		\frac{1}{90} - \frac{x}{90^2}, & x\in [0\degree, 90\degree] \\
		0, & x\not\in \Gamma_\text{tri}^{i}
		\end{array}
		\right. \right \}_{i=1}^{9}$\\
		\hline
	\end{tabu}
	\caption{List of input uncertainty data for all scenarios (scenario index: $j = 1,\dots,6$).}
	\label{table:scenario_pdfs}
\end{table}

For each base layout, we consider the 6 deviation scenarios and use stochastic collocation to calculate an interpolant of $T_{avg}$ as a function of the support of all 18 random variables belonging to that scenario. For a specific scenario, the collocation points generated by the algorithm are joint realizations of all 18 variables. Each realization encodes an array geometry, in which the 9 BHEs are individually deviated. For those array geometries, we simulate the extraction regime detailed above and calculate $T_{avg}$ as an exact solution for that node.

Additionally, because of the nature of stochastic collocation providing an interpolant, it is possible to consider another variation for each scenario where instead of uniformly distributed directions, we assume them to be triangularly distributed. This can be done simply by resampling the interpolant and does not require the solution of any additional PDEs. The downside of this is the fact that the error indicators of the collocation method only apply to the interpolant with regards to the original uniform input, however tests have shown that the indicators remain mostly reliable for triangular distributions. Theoretically, this can be applied to all input distributions to generate additional information for free. However, caution has to be exercised if the new distributions differ significantly. Formally, we consider triangularly distributed directions of deviation by introducing the set of random variables $\left \{ {\mathcal{X}}_\text{tri}^{i} \right \}_{i=1}^{9}$ as specified in Table \ref{table:scenario_pdfs} and Figure \ref{fig:pdf-input} at the top right.

\subsection{Results}

\begin{figure}[t!]
	%\begin{figure}[htbp]
	\centering
	\begin{subfigure}[b]{0.58\textwidth}
		\centering
		\includegraphics[width=\textwidth]{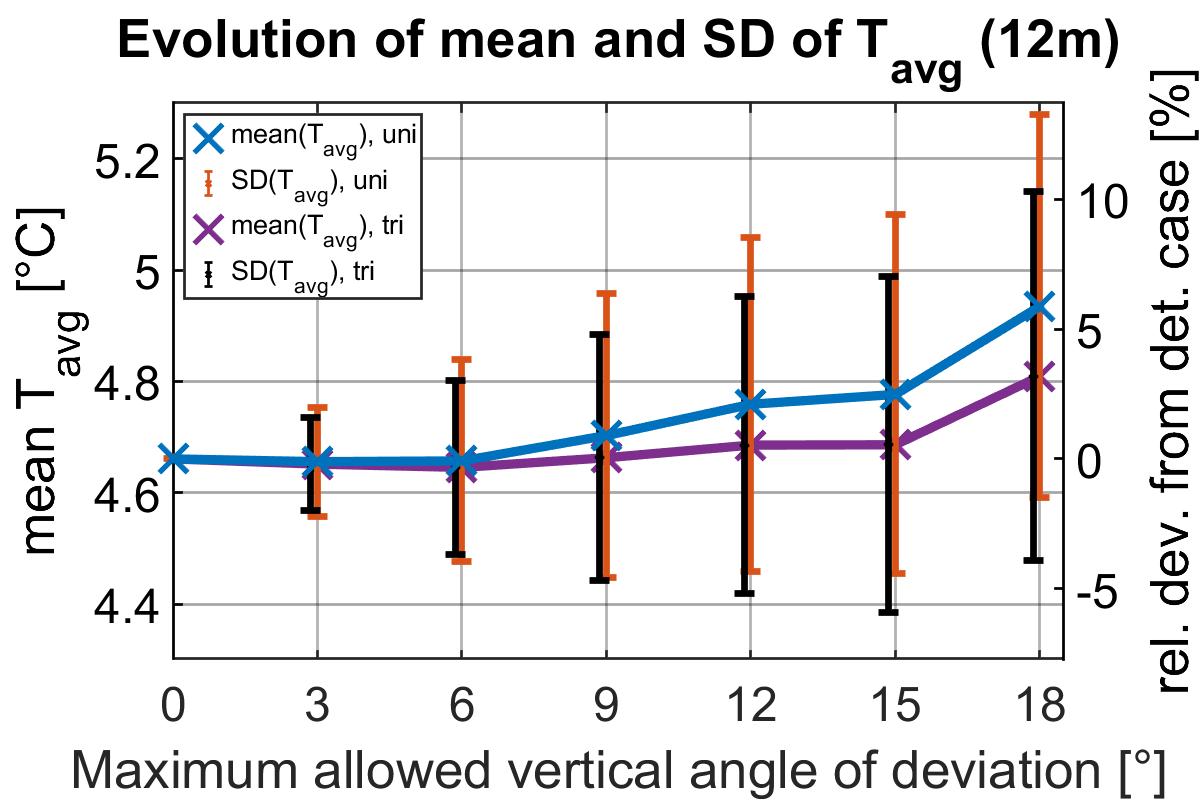}
		%\caption{Layout and placement of the BHEs in the BTES.}
		%{{\small Network 1}}
		%\label{fig:mean and std of net14}
	\end{subfigure}
	\vskip\baselineskip
	%\hspace{4mm}
	\begin{subfigure}[b]{0.58\textwidth}
		\centering
		\includegraphics[width=\textwidth]{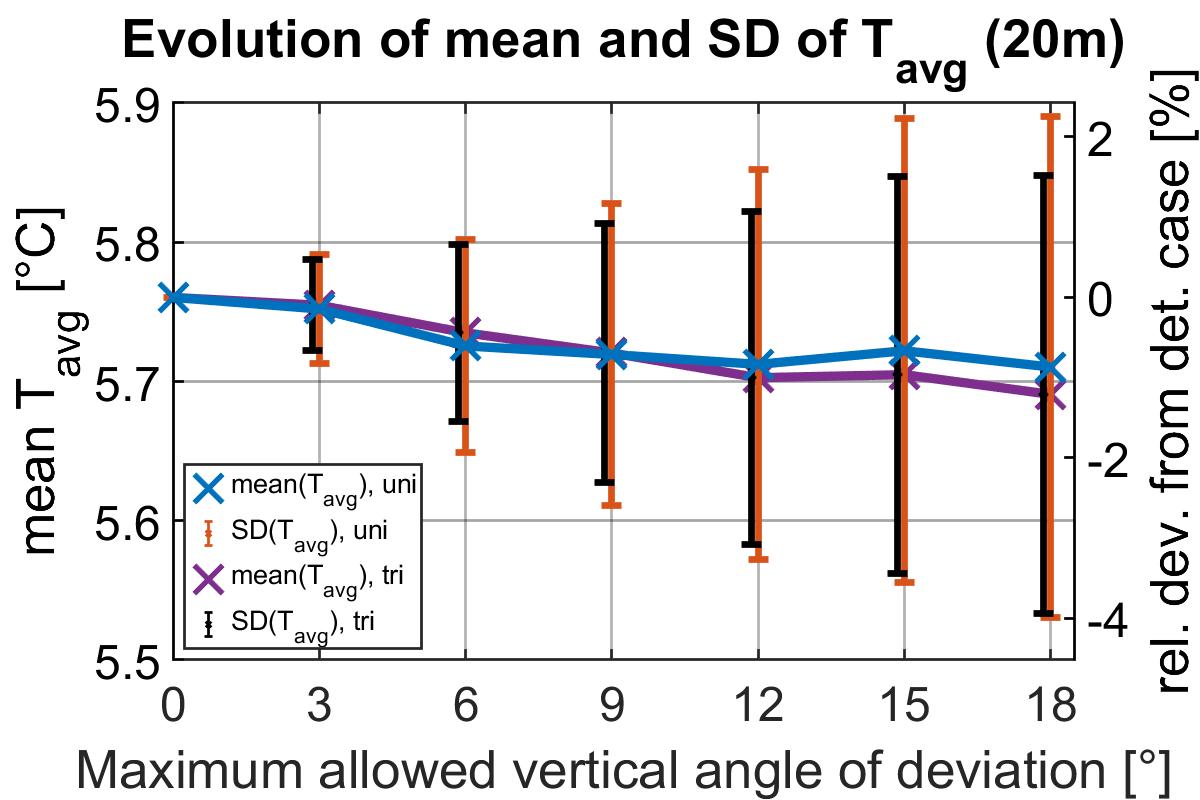}
		%\caption{Layout and placement of the BHEs in the BTES.}
		%{{\small Network 2}}
		%\label{fig:mean and std of net24}
	\end{subfigure}
	\vskip\baselineskip
	\centering
	\begin{subfigure}[b]{0.58\textwidth}
		\centering
		\includegraphics[width=\textwidth]{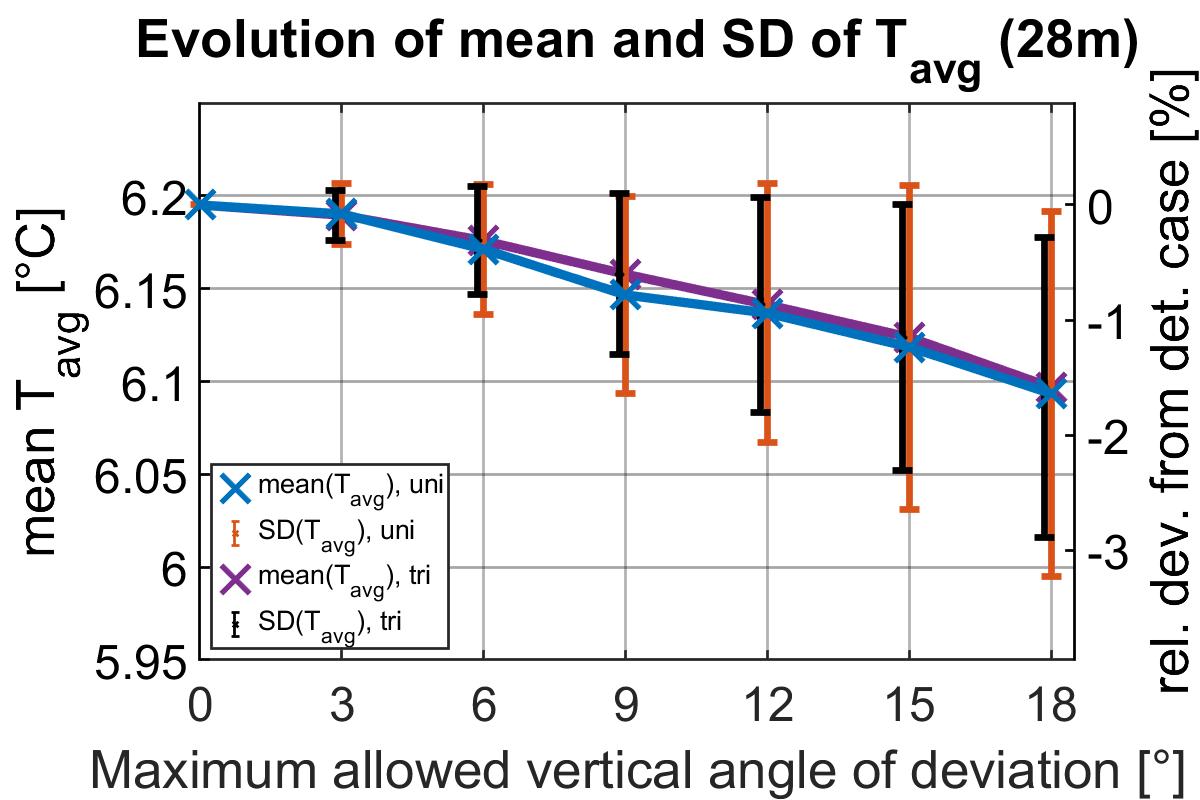}
		%\caption{Layout and placement of the BHEs in the BTES.}
		%{{\small Network 2}}
		%\label{fig:mean and std of net24}
	\end{subfigure}
	\caption{Evolution of the calculated means and standard deviations of $T_{avg}$ for the three base layouts across all 6 scenarios using uniformly distributed direction angles, as well as the reevaluation using triangular distributions.}
	\label{mean_evo}
\end{figure}

\begin{figure}[htbp]
	%\begin{figure}[htbp]
	\centering
	\begin{subfigure}[b]{0.58\textwidth}
		\centering
		\includegraphics[width=\textwidth]{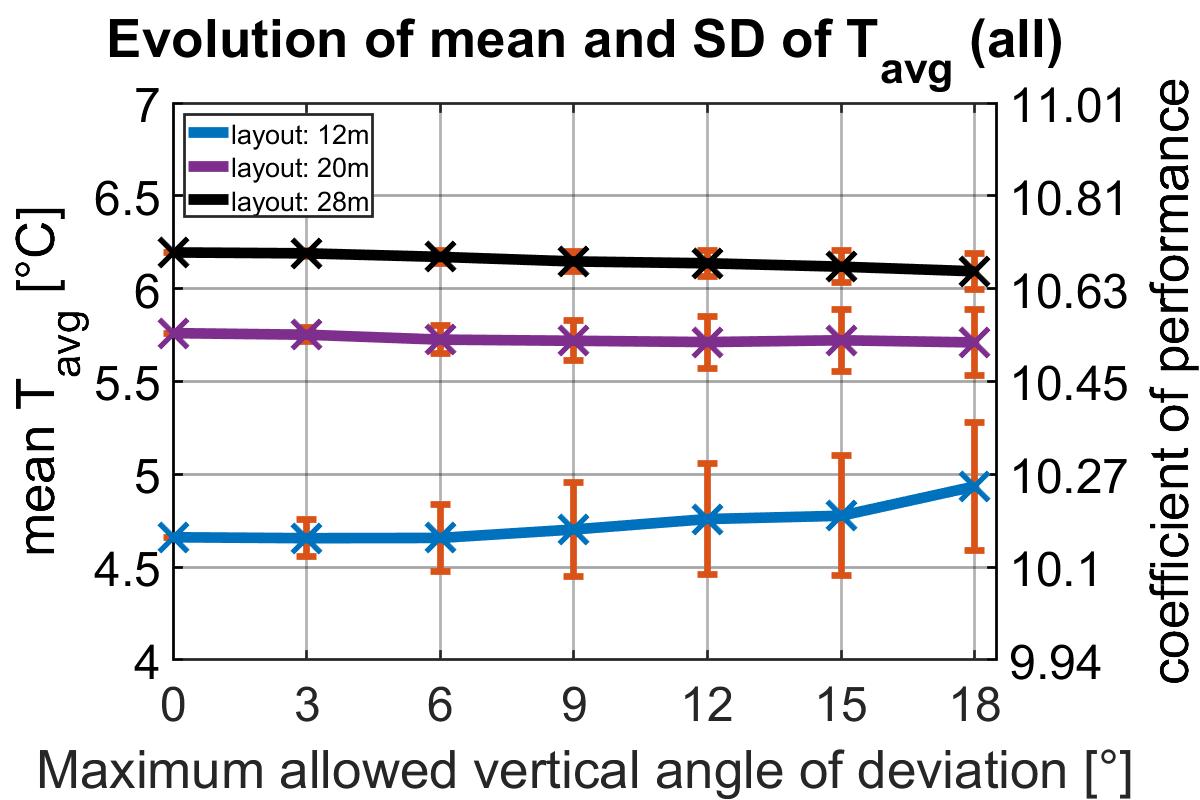}
		%\caption{Layout and placement of the BHEs in the BTES.}
		%{{\small Network 1}}
		%\label{fig:mean and std of net14}
	\end{subfigure}
	%\hspace{4mm}
	\vskip\baselineskip
	\begin{subfigure}[b]{0.58\textwidth}
		\centering
		\includegraphics[width=\textwidth]{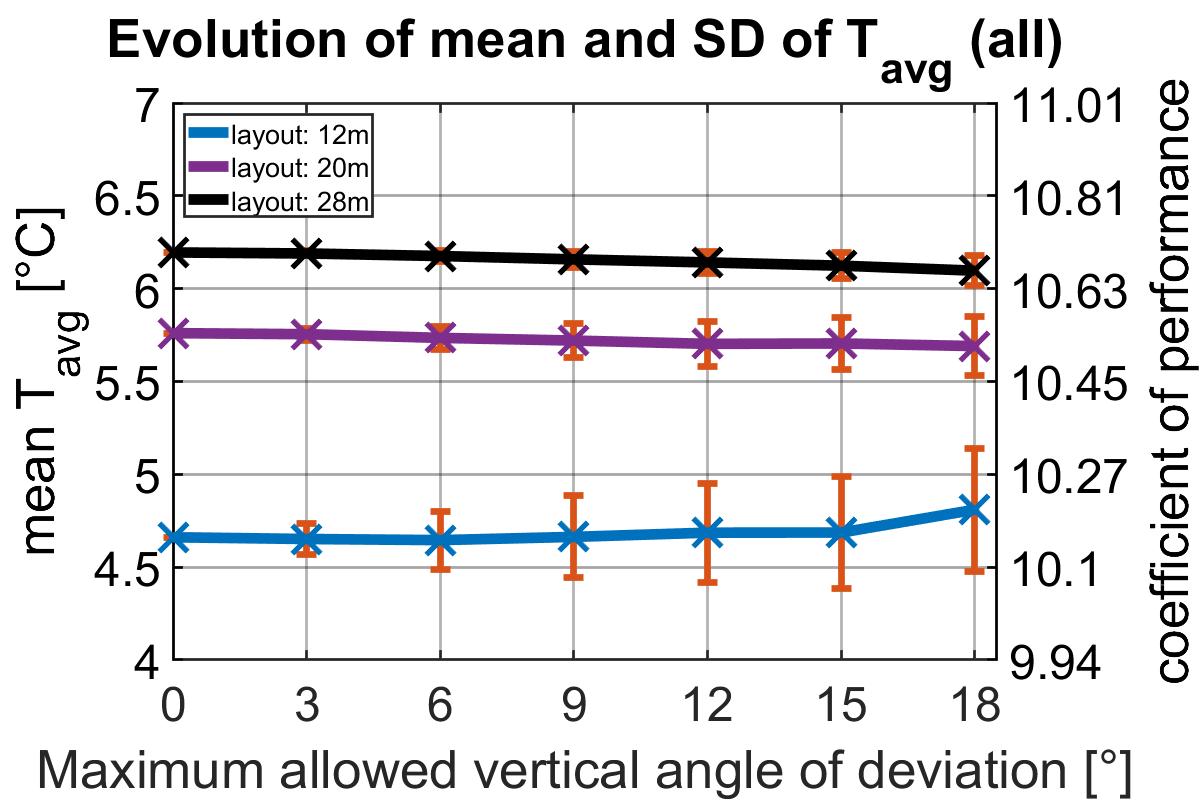}
		%\caption{Layout and placement of the BHEs in the BTES.}
		%{{\small Network 2}}
		%\label{fig:mean and std of net24}
	\end{subfigure}
	\caption{Evolution of the calculated means and standard deviations of $T_{avg}$ and the accompanying COP for the three base layouts across all 6 scenarios in comparison for uniformly (left) and triangularly (right) distributed direction angles.}
	\label{mean_evo_all}
\end{figure}

Having calculated the interpolant for all 6 scenarios for each base layout, we estimate the mean and standard deviation of the quantity of interest, in this case $T_{avg}$, for each one. Figure \ref{mean_evo} and \ref{mean_evo_all} show the evolution of these quantities for each layout across the successive scenarios for both uniformly and triangularly distributed angles of direction of the deviations. Additionally, Figure \ref{mean_evo} tracks the relative deviation from the deterministic case (the reference layout without deviations), while Figure \ref{mean_evo_all} compares the coefficient of performance.

\begin{figure}[t!]
	%\begin{figure}[htbp]
	\centering
	\begin{subfigure}[b]{0.40\textwidth}
		\centering
		\includegraphics[width=\textwidth]{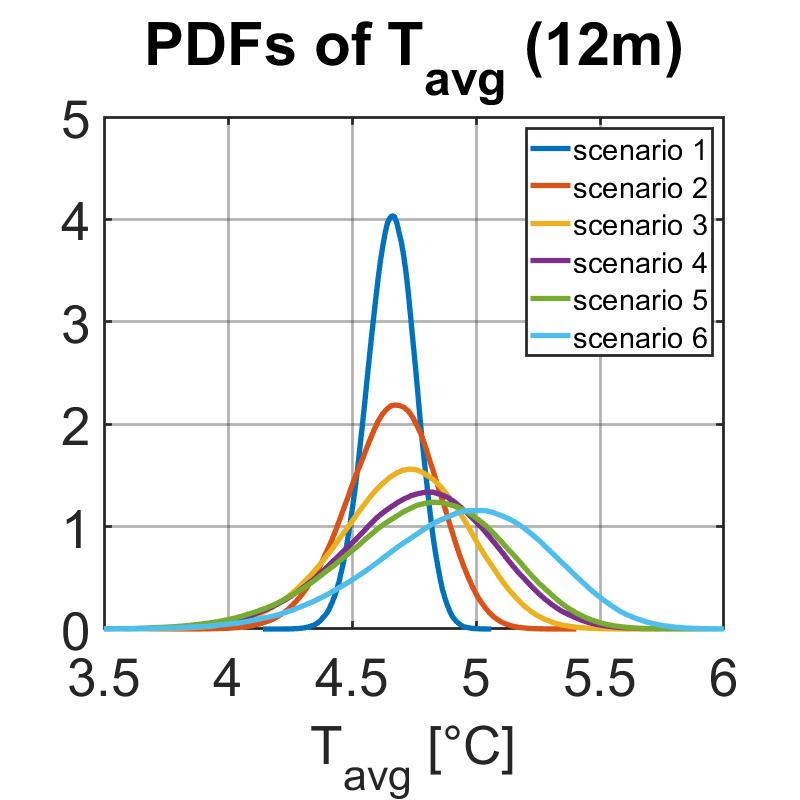}
		%\caption[Network2]%
		%{{\small Network 1}}
		%\label{fig:mean and std of net14}
	\end{subfigure}
	\hspace{8mm}
	\begin{subfigure}[b]{0.40\textwidth}
		\centering
		\includegraphics[width=\textwidth]{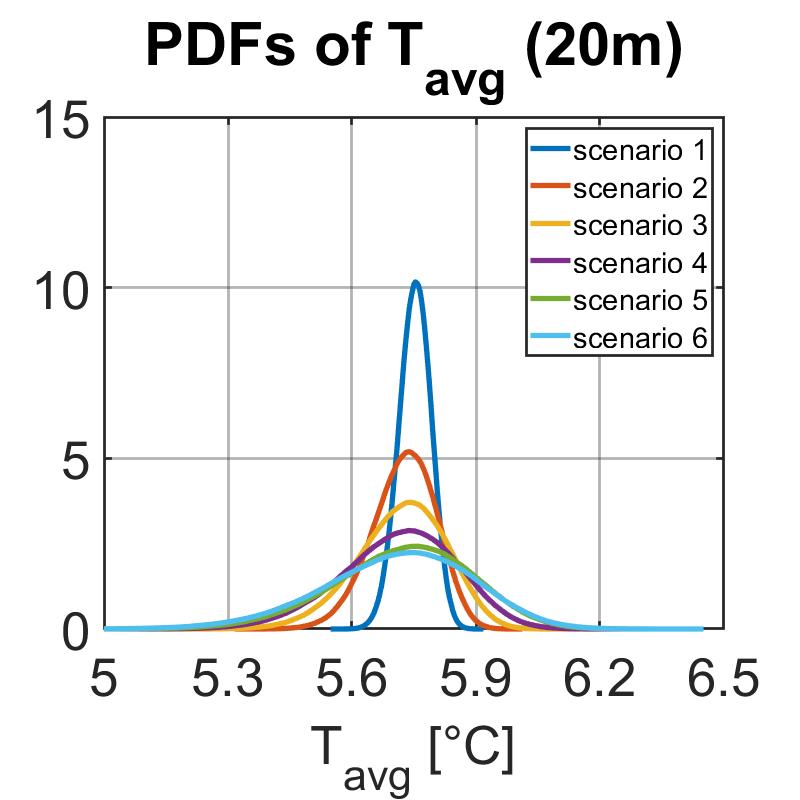}
		%\caption[]%
		%{{\small Network 2}}
		%\label{fig:mean and std of net24}
	\end{subfigure}
	\vskip\baselineskip
	\begin{subfigure}[b]{0.40\textwidth}
		\centering
		\includegraphics[width=\textwidth]{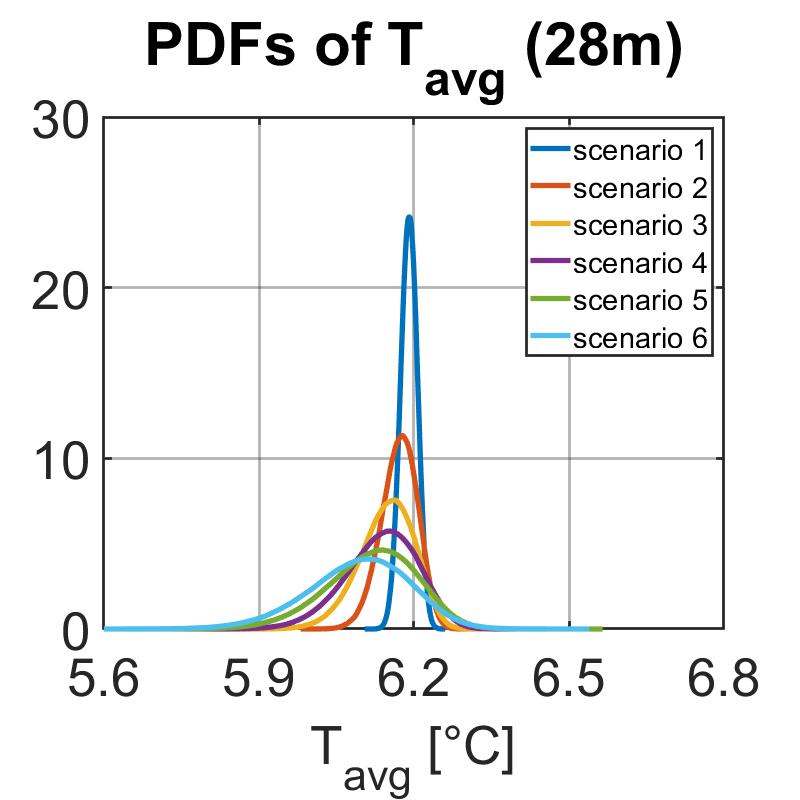}
		%\caption[]%
		%{{\small Network 2}}
		%\label{fig:mean and std of net24}
	\end{subfigure}
	\hspace{8mm}
	\begin{subfigure}[b]{0.40\textwidth}
		\centering
		\includegraphics[width=\textwidth]{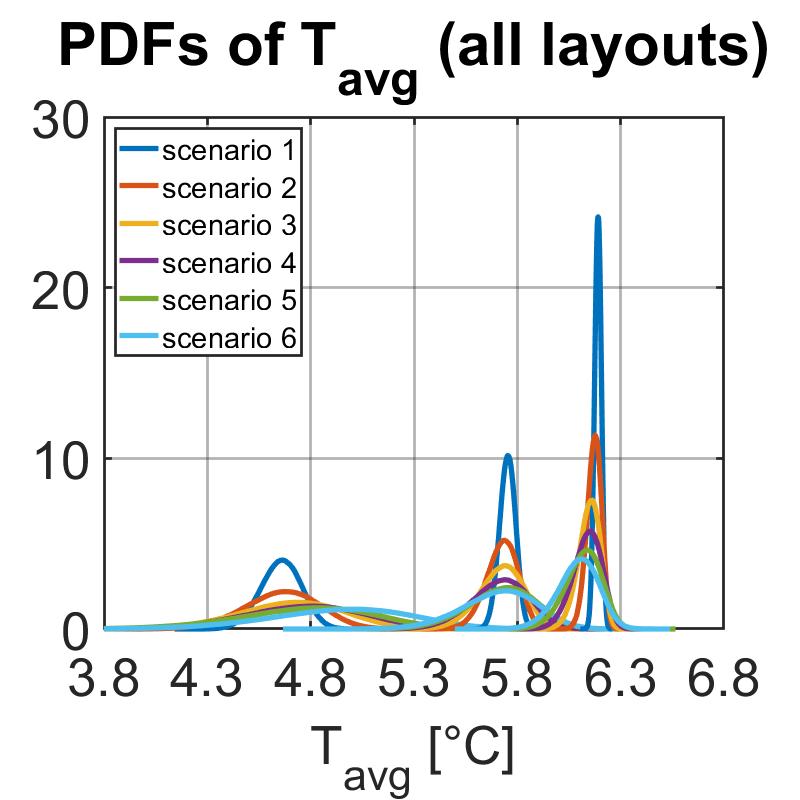}
		%\caption[]%
		%{{\small Network 2}}
		%\label{fig:mean and std of net24}
	\end{subfigure}
	\caption{Probability density functions of $T_{avg}$ for all base layouts and scenarios using uniformly distributed directions of deviations.}
	\label{fig:pdf1-6_uni}
\end{figure}

\begin{figure}[t!]
	%\begin{figure}[htbp]
	\centering
	\begin{subfigure}[b]{0.40\textwidth}
		\centering
		\includegraphics[width=\textwidth]{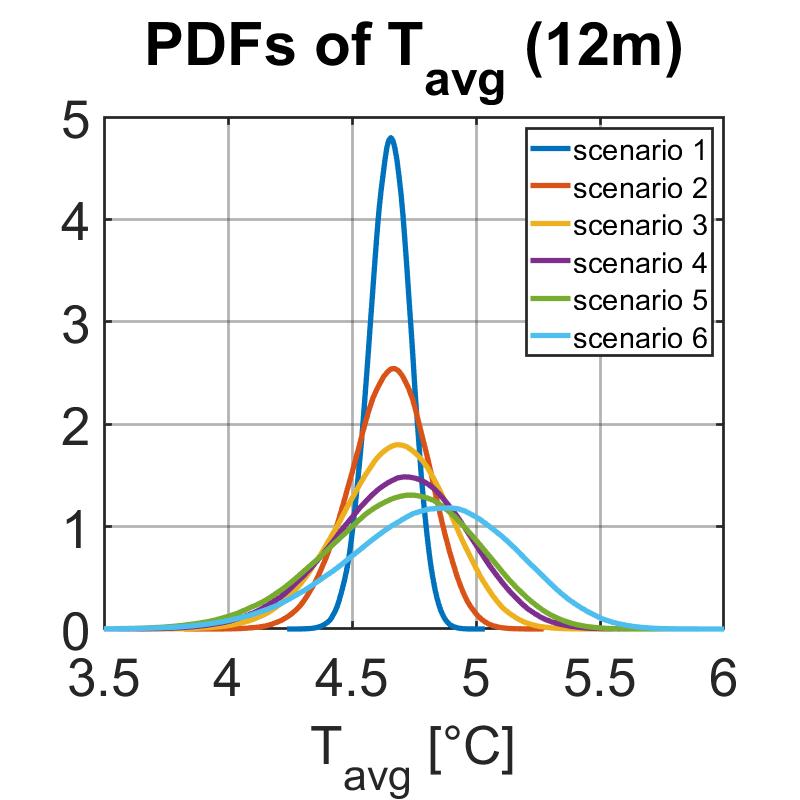}
		%\caption[Network2]%
		%{{\small Network 1}}
		%\label{fig:mean and std of net14}
	\end{subfigure}
	\hspace{8mm}
	\begin{subfigure}[b]{0.40\textwidth}
		\centering
		\includegraphics[width=\textwidth]{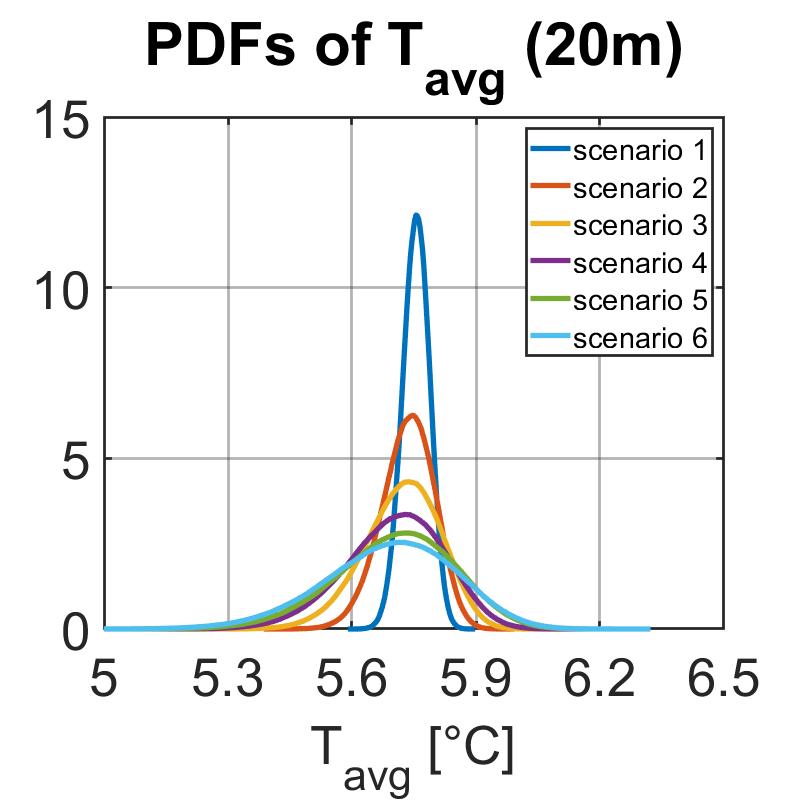}
		%\caption[]%
		%{{\small Network 2}}
		%\label{fig:mean and std of net24}
	\end{subfigure}
	\vskip\baselineskip
	\begin{subfigure}[b]{0.40\textwidth}
		\centering
		\includegraphics[width=\textwidth]{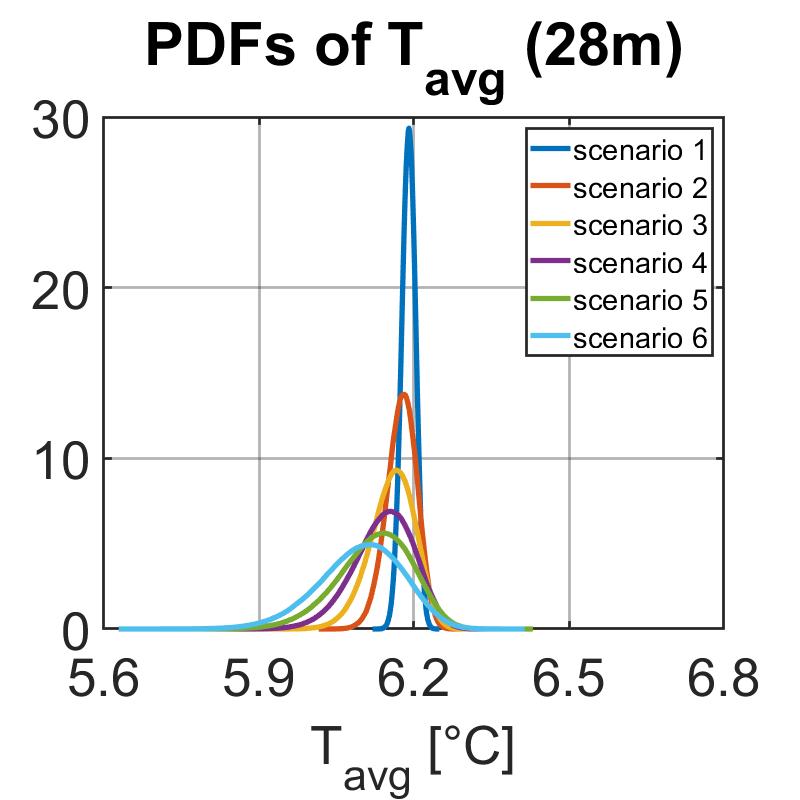}
		%\caption[]%
		%{{\small Network 2}}
		%\label{fig:mean and std of net24}
	\end{subfigure}
	\hspace{8mm}
	\begin{subfigure}[b]{0.40\textwidth}
		\centering
		\includegraphics[width=\textwidth]{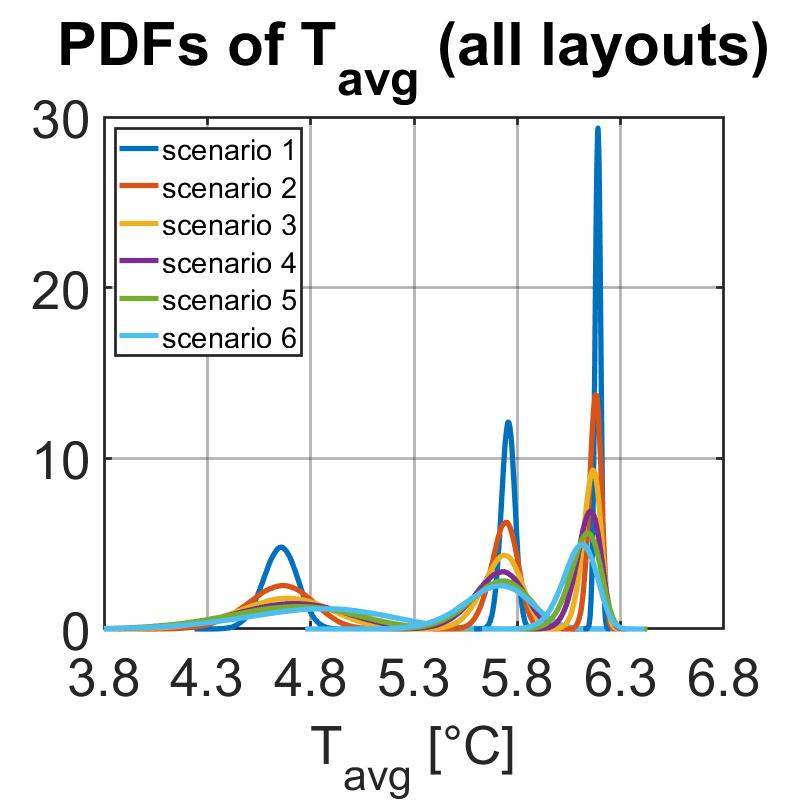}
		%\caption[]%
		%{{\small Network 2}}
		%\label{fig:mean and std of net24}
	\end{subfigure}
	\caption{Probability density functions of $T_{avg}$ for all base layouts and scenarios using triangularly distributed directions of deviations.}
	\label{fig:pdf1-6_tri}
\end{figure}

The means and standard deviations of the outlet temperature could be also estimated using Monte Carlo sampling. However, having access to the interpolant through stochastic collocation provides much more statistical information. As mentioned in section \ref{sec:Statistics}, the interpolant can be sampled with realizations of the random variables. Performing a kernel density estimation on the results provides an estimate of the pdf $\rho_{T_{avg}}$ of $T_{avg}$ itself for each scenario. The pdfs for all layouts and scenarios are shown in Figures \ref{fig:pdf1-6_uni} and \ref{fig:pdf1-6_tri}.


\begin{figure}[t!]
	%\begin{figure}[htbp]
	\centering
	\begin{subfigure}[b]{0.32\textwidth}
		\centering
		\includegraphics[width=\textwidth]{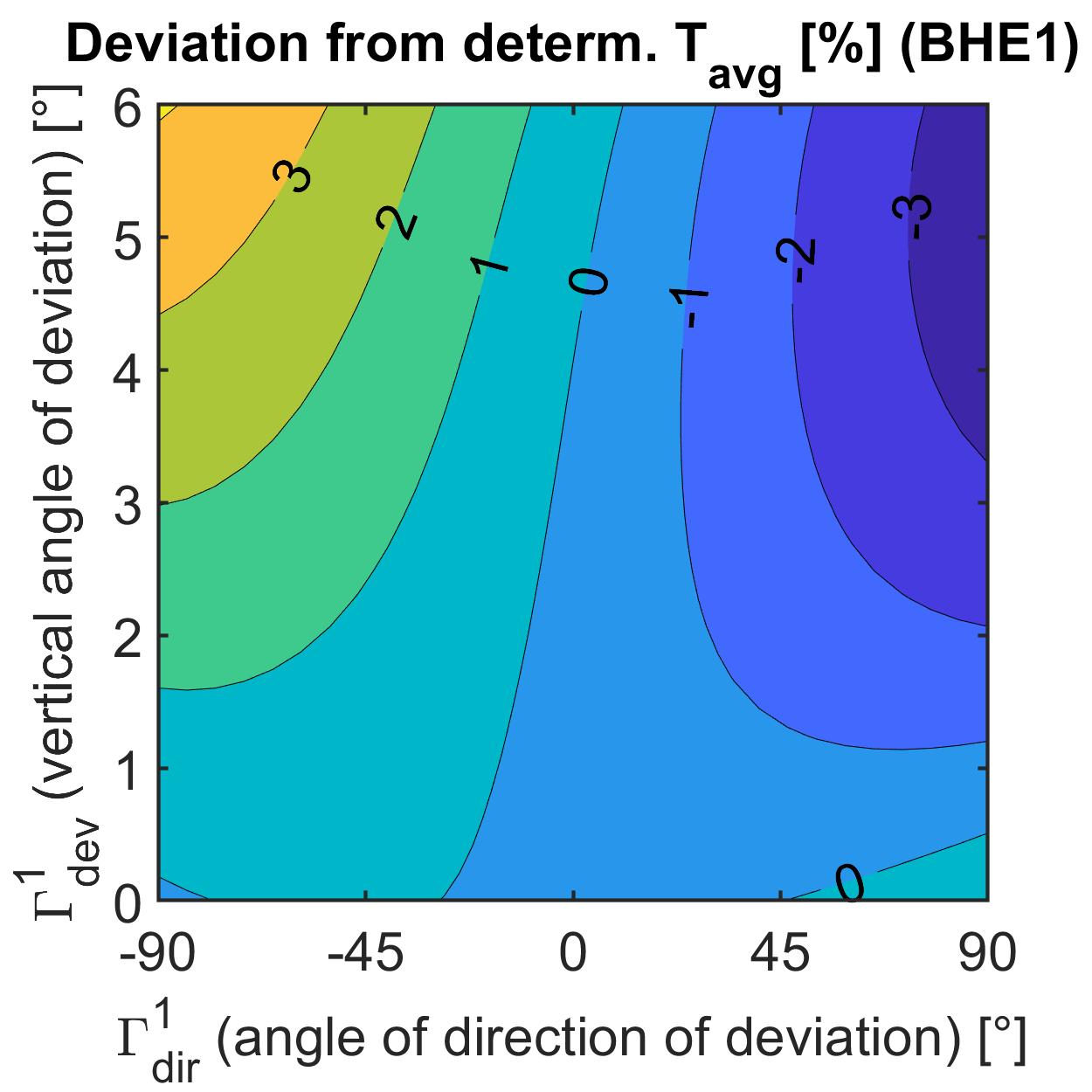}
		%\caption[Network2]%
		%{{\small Network 1}}
		%\label{fig:mean and std of net14}
	\end{subfigure}
	\hfill
	\begin{subfigure}[b]{0.32\textwidth}
		\centering
		\includegraphics[width=\textwidth]{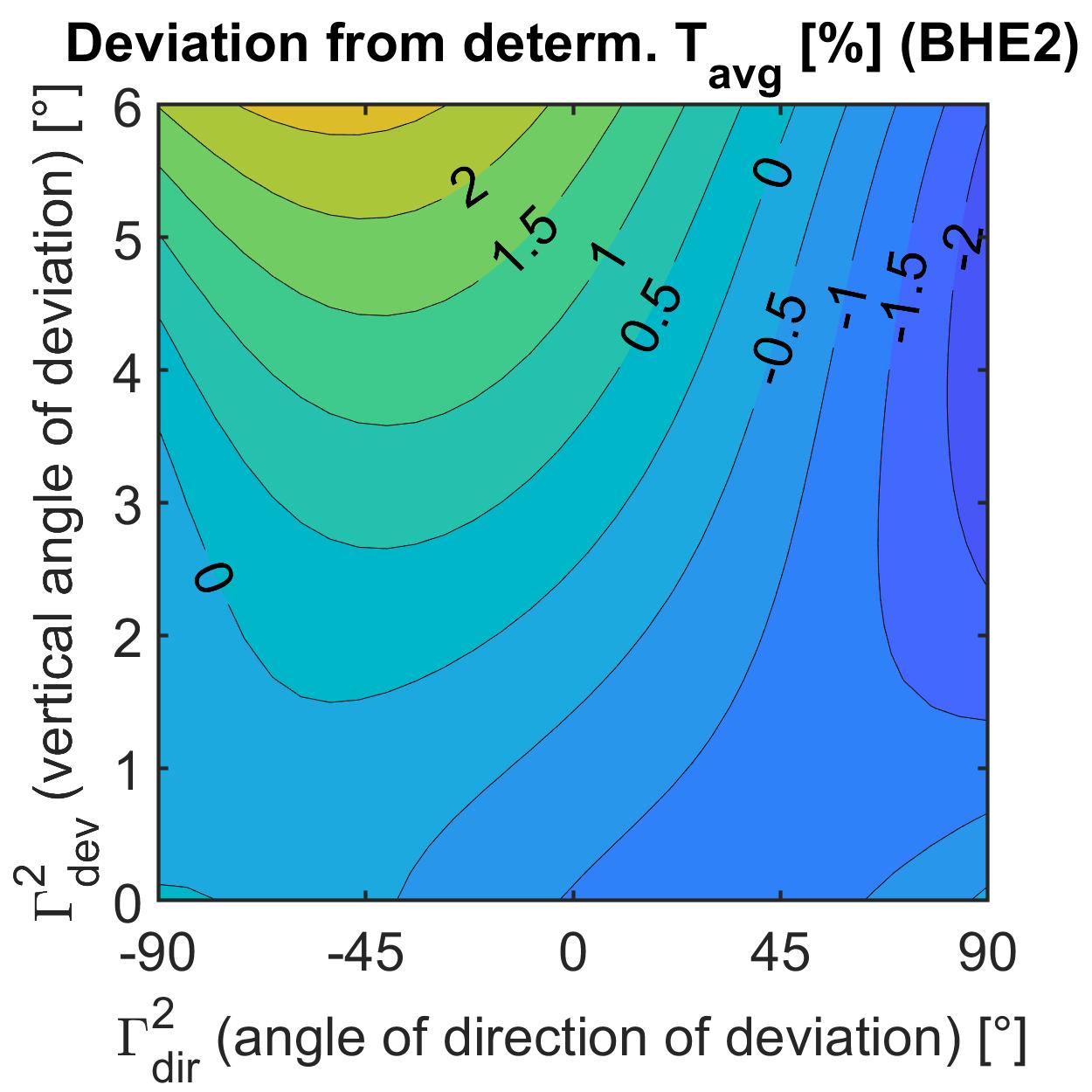}
		%\caption[]%
		%{{\small Network 2}}
		%\label{fig:mean and std of net24}
	\end{subfigure}
	\hfill
	\begin{subfigure}[b]{0.32\textwidth}
		\centering
		\includegraphics[width=\textwidth]{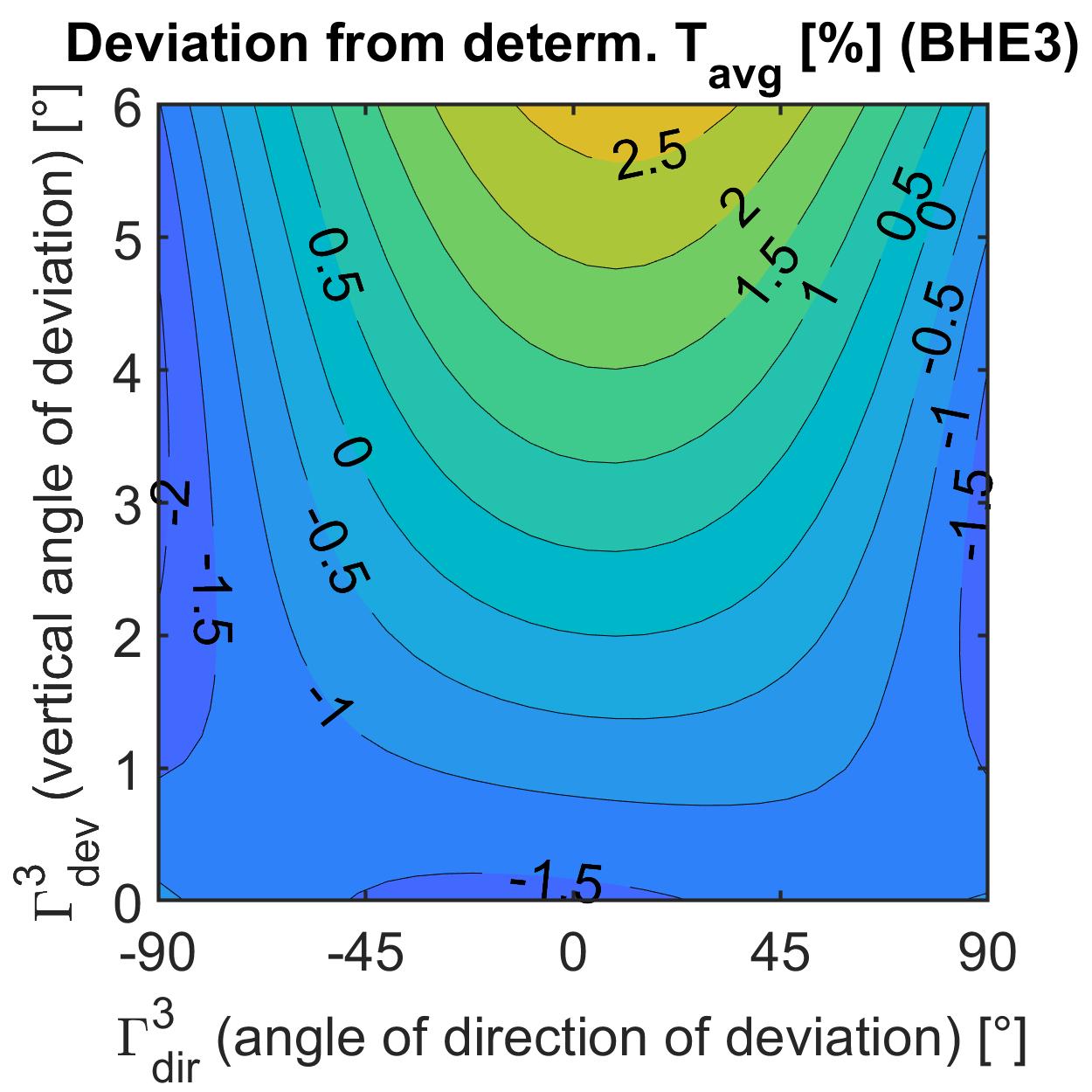}
		%\caption[]%
		%{{\small Network 2}}
		%\label{fig:mean and std of net24}
	\end{subfigure}
	\vskip\baselineskip
	\begin{subfigure}[b]{0.32\textwidth}
		\centering
		\includegraphics[width=\textwidth]{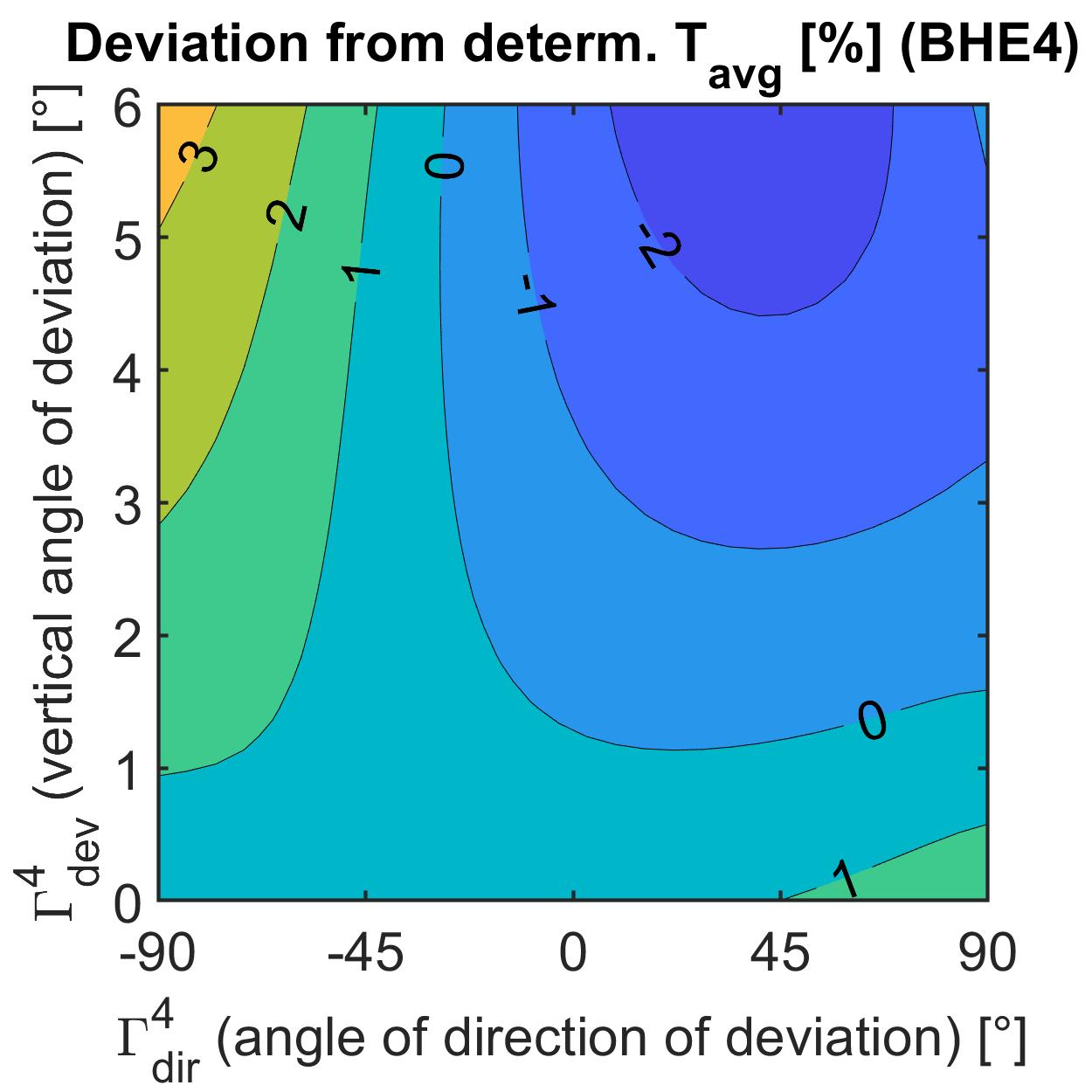}
		%\caption[]%
		%{{\small Network 2}}
		%\label{fig:mean and std of net24}
	\end{subfigure}
	\hfill
	\begin{subfigure}[b]{0.32\textwidth}
		\centering
		\includegraphics[width=\textwidth]{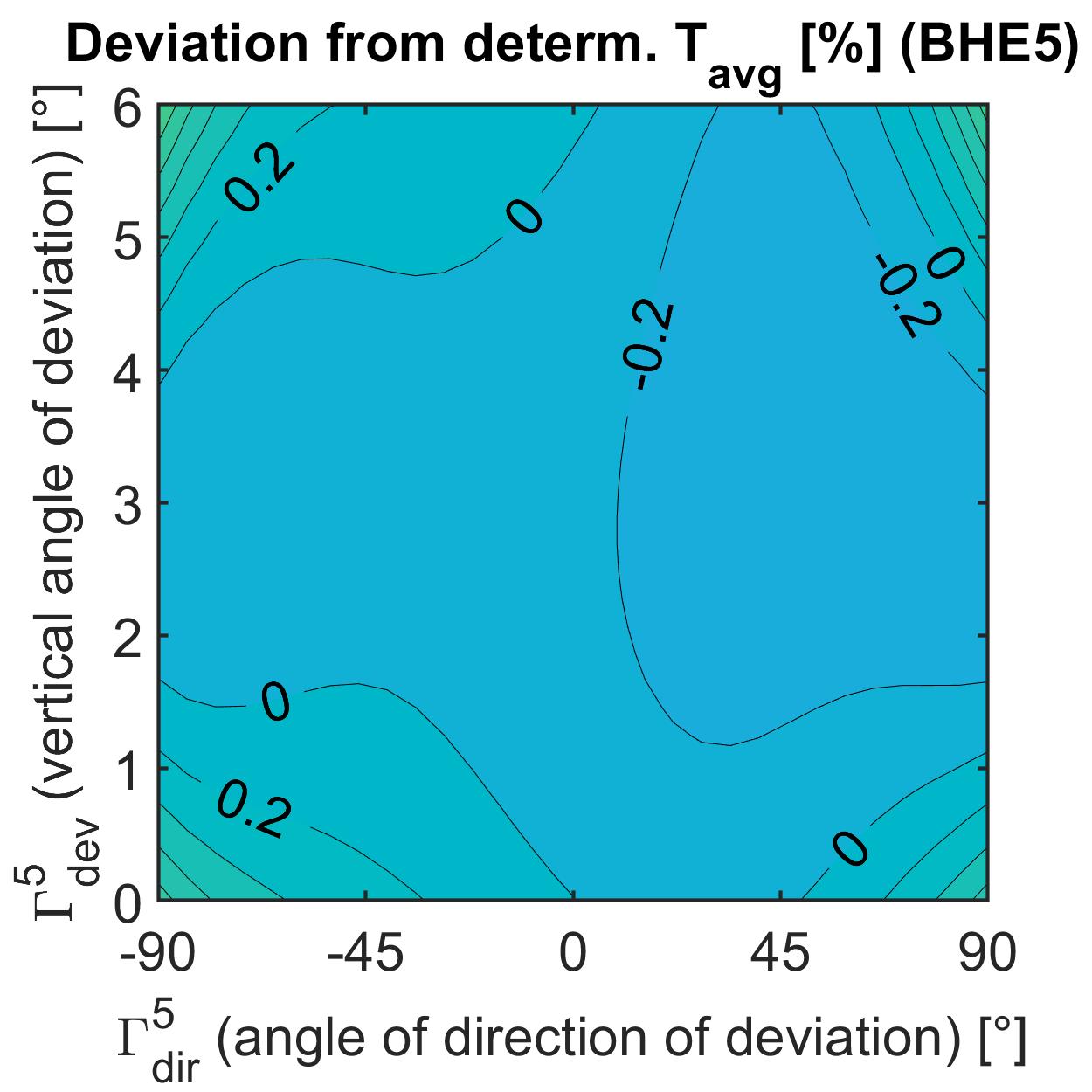}
		%\caption[]%
		%{{\small Network 2}}
		%\label{fig:mean and std of net24}
	\end{subfigure}
	\hfill
	\begin{subfigure}[b]{0.32\textwidth}
		\centering
		\includegraphics[width=\textwidth]{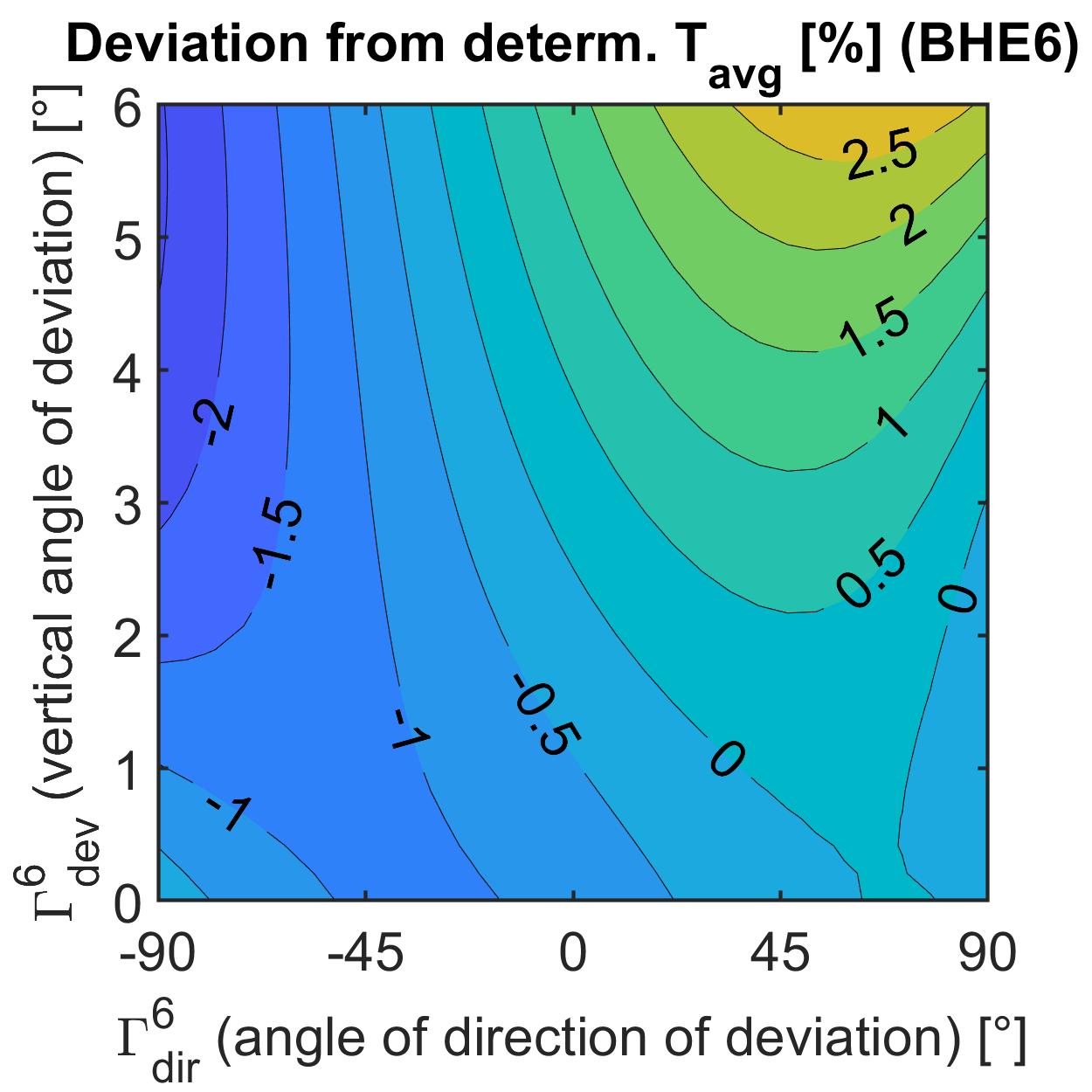}
		%\caption[]%
		%{{\small Network 2}}
		%\label{fig:mean and std of net24}
	\end{subfigure}
	\vskip\baselineskip
	\begin{subfigure}[b]{0.32\textwidth}
		\centering
		\includegraphics[width=\textwidth]{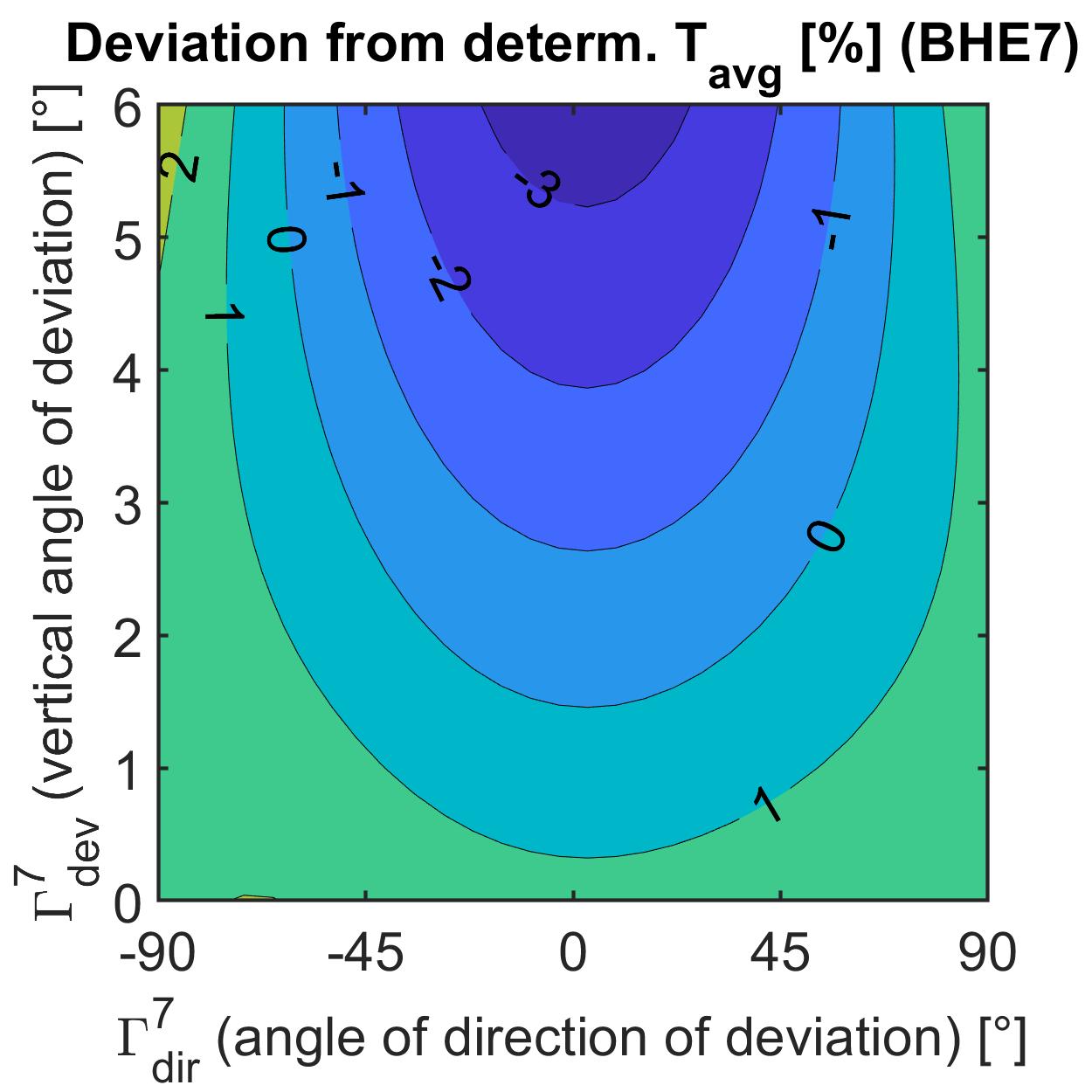}
		%\caption[]%
		%{{\small Network 2}}
		%\label{fig:mean and std of net24}
	\end{subfigure}
	\hfill
	\begin{subfigure}[b]{0.32\textwidth}
		\centering
		\includegraphics[width=\textwidth]{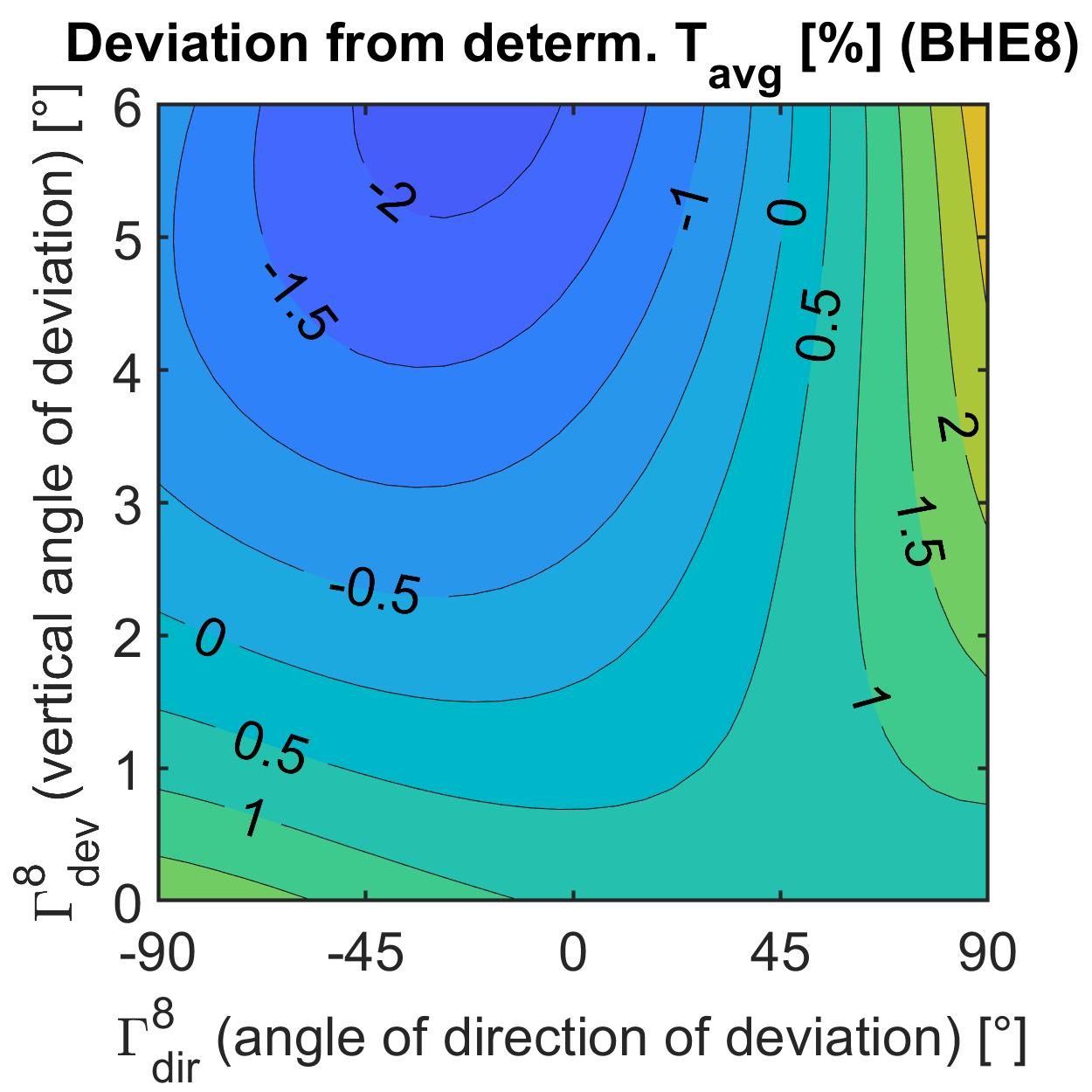}
		%\caption[]%
		%{{\small Network 2}}
		%\label{fig:mean and std of net24}
	\end{subfigure}
	\hfill
	\begin{subfigure}[b]{0.32\textwidth}
		\centering
		\includegraphics[width=\textwidth]{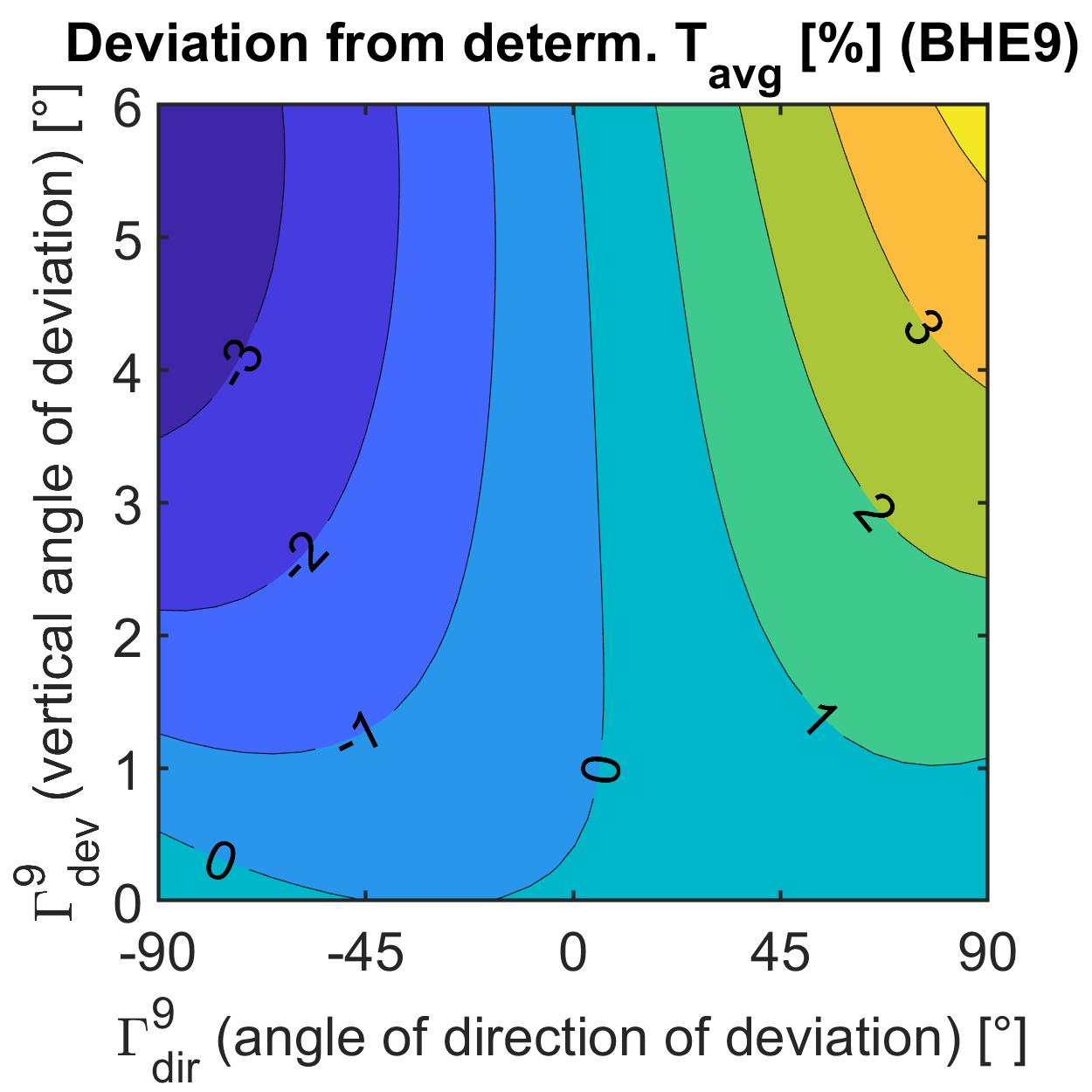}
		%\caption[]%
		%{{\small Network 2}}
		%\label{fig:mean and std of net24}
	\end{subfigure}
	\caption{Plots of the partially integrated interpolant of $T_{avg}$ for each pair of random variables responsible for a BHE. The plots are arranged to conform to the layout of the actual facility as shown in Figure \ref{BTES_layout}. The interpolant of the 12m-layout in the second scenario is used.}
	\label{fig:BHE1-9}
\end{figure}

Lastly, we illustrate the interpolant of $T_{avg}$ for the 12m-layout in the second scenario, corresponding to an allowed vertical angle of up to $6\degree$. For each BHE, we evaluated the interpolant on the support $\Gamma_\text{dir}^{i}$ and $\Gamma_\text{dev}^{i}$ of the 2 random variables $\mathcal{X}_\text{dir}^{i}$ and $\mathcal{X}_\text{dev}^{i}$ responsible for the deviations of that BHE. For every other random variable, we integrate over its support, eliminating that variable. This allows us to investigate the effects of the deviations of a single BHE and plot the quantity of interest $T_{avg}$ as a function of the direction of deviation and vertical angle of deviation of a particular BHE. Instead of the absolute temperature of $T_{avg}$, we plotted the percentual deviation from the reference case's $T_{avg}$, meaning the $T_{avg}$ of the original, planned layout. This allows us to investigate how the performance of the BHE array is improved or worsened by the deviations of that BHE. The resulting 9 2D-plots are shown in Figure \ref{fig:BHE1-9}.

\section{Evaluation and discussion}
\label{sec:discussion}

Based on the results illustrated in the previous section in Figures \ref{mean_evo} and \ref{mean_evo_all}, some general trends can be observed for the estimated means of $T_{avg}$. For the 12m-layout, the mean performance improves with the maximum allowed inclination.
For the 20m-layout, a minuscule decrease in mean performance can be observed with increasing maximum inclination. Also for the 28m-layout the mean $T_{avg}$ decreases with maximum allowed inclination, but in contrast to the 20m-layout, the effect is significantly stronger.

These differing observations can be explained by the spacing of neighboring BHEs of the respective layouts' reference cases. The 12m-layout reference case supplies only a $T_{avg}$ of approximately $4.7~\celsius$ whereas the 28m-layout reference case operates at almost $6.2~\celsius$. Due the close spacing, it can be assumed that in the 12m-layout there is significant thermal interference between the BHEs leading to the lower $T_{avg}$. Therefore, a deviation from a perpendicular bore path is more likely to increase the spacing between BHEs at depth alleviating thermal interference and subsequently improving the average BHE performance than to decrease the distance between BHEs and impair the performance even further.

In the 28m-layout, spacing between neighboring BHEs is increased by 40\% and the results show that, on average, bore path deviations will reduce the array's performance. Even when taking into account the indicated standard deviations, an improvement of $T_{avg}$ is very unlikely. Consequently, it can be assumed that in the 28m-layout reference case the spacing is large enough to nullify any thermal interaction between BHEs. Deviations from the perpendicular bore path can only reduce the array's performance by decreasing the distance between BHEs at depth bringing them into the range of thermal interference, whereas deviations that increase BHE spacing have no beneficial effect.

%% Die folgenden beiden Absätze müssen vllt verschoben werden. Irgendwie sind sie ein bisschen out of place. Vllt hast Du eine Idee. Ansonsten schaue ich mir das morgen nochmal mit frischem Kopf an.
Typically, subsurface anisotropy like bedding or foliation causes bore paths to veer off their planned trajectory. As a result, azimuth and inclination of bore path deviations can be expected to be normally distributed around a specific direction. In contrast to that, uniform distributions of bore path deviations are rather unrealistic, but can be used to analyse worst case scenarios where extreme deviations are just as likely as slight ones.
Since the results for a triangular distribution can be obtained without additional effort from the uniform distribution, it is used to demonstrate the impact of the distribution type. While there is no discernible difference between the distribution types in the 20m-layout and the 28m-layout, a marked difference can be observed in the 12m-layout: Uniformly distributed bore path deviations benefit more than triangularly distributed ones, especially for maximum inclination angles of more than $12\degree$.

The same rational can be applied as before: in the closely spaced 12m-layout, the BHEs impair each other due to thermal interference. While a uniform distribution can allow for divergent bore paths that lead to larger spacing at depth, a triangular distribution tends to keep the bore paths in 'bunched up' array geometries, where individual BHEs stay closer to each other. In the 28-m layout, there is no thermal interference to begin with. Consequently, diverging borehole paths cannot compensate any impairment and the difference in distributions types is negligible. This effect will be explained in more detail later in this section.

Comparison of the different layouts shows, that standard deviation grows with decreasing spacing between the BHEs (Figure \ref{mean_evo_all}). This means that it becomes more likely that borehole path uncertainty will affect the BHE array performance, the closer the BHE spacing gets.
Since the standard deviation of the inputs' uncertainty increases with each scenario, the estimated standard deviations of $T_{avg}$ increases, as well, for both the uniformly (Figure \ref{fig:pdf1-6_uni}) and triangularly (Figure \ref{fig:pdf1-6_tri}) distributed variables. The uniform distribution's standard deviation of $T_{avg}$ is consistently larger than its triangular counterpart, which is also expected since the more likely deviations of the triangular distribution are geometrically similar and thus perform similarly as well.

Taking into account both the mean and standard deviation of $T_{avg}$, $\overline{T_{avg}} \pm SD(T_{avg})$  can be compared to the respective reference case (without bore path deviation). The maximum decrease/increase ranges from -6\% to 13.3\% for the 12m-layout, from -4\% to 2.3\% for the 20m-layout and from -3,3\% to 0.2\% in the 28m-layout across all scenarios. This information can already serve as a general idea of how much the deviations will affect performance statistically, but this analysis can be further refined by using of the calculated pdfs $\rho_{T_{avg}}$ of $T_{avg}$ in Figures \ref{fig:pdf1-6_uni} and \ref{fig:pdf1-6_tri}. By numerically solving the one-sided $95\%$-confidence intervals
\begin{align}
	\begin{aligned}
	P(T_{avg} > T_\text{min}) &= 95\% \\
	\Leftrightarrow \int_{T_\text{min}}^{\infty} \rho_{T_{avg}}(x) \ dx &= 95\%
	\end{aligned}
\end{align}
for all scenarios, a $T_\text{min}$ value can be extracted for each one. $T_\text{min}$ represents a lower bound of performance in the sense that 95\% of all possible geometries in a given scenario will perform at least as well as $T_\text{min}$. It is calculated for each layout and for all scenarios so that the relative difference to the deterministic $T_{avg}$ of each layout's reference case can be determined as a 'worst-case'. The results are listed in Table \ref{table:T_min}.
\begin{table}[htbp]
	\centering
	\begin{tabu}{|l|[2pt]l|l|l|l|l|l|l|}
		\hline
		\multicolumn{8}{|c|}{uniformly distributed directions} \\
		\hline
		layout& det. $T_{avg}$ & $3\degree$ & $6\degree$ & $9\degree$ & $12\degree$ & $15\degree$ & $18\degree$  \\
		\tabucline[2pt]{-}
		12m & $4.66~\celsius$ & -3.67\% &   -6.78\% &   -8.54\% &   -8.97\% &   -9.61\% &   -7.04\% \\
		\hline
		20m & $5.76~\celsius$ & -1.28\% &   -2.92\% &   -4.01\% &   -5.09\% &   -5.71\% &   -6.32\% \\
		\hline
		28m & $6.20~\celsius$ & -0.52\% &   -1.37\% &   -2.30\% &   -2.93\% &   -3.72\% &   -4.41\% \\
		\hline
		\multicolumn{8}{|c|}{triangularly distributed directions} \\
		\hline
		layout& det. $T_{avg}$ & $3\degree$ & $6\degree$ & $9\degree$ & $12\degree$ & $15\degree$ & $18\degree$  \\
		\tabucline[2pt]{-}
		12m & $4.66~\celsius$ & -3.20\% &   -6.09\% &   -8.09\% &   -9.30\% & -10.69\% &   -9.15\%  \\
		\hline
		20m & $5.76~\celsius$ & -1.06\% &   -2.36\% &   -3.50\% &   -4.63\% &   -5.26\% &   -5.97\%  \\
		\hline
		28m & $6.20~\celsius$ & -0.46\% &   -1.12\% &   -1.83\% &   -2.52\% &   -3.20\% &   -3.87\%  \\
		\hline
	\end{tabu}
	\caption{Relative difference between $T_\text{min}$ and the deterministic $T_{avg}$ for all layouts and scenarios.}
	\label{table:T_min}
\end{table}

Considering the worst-case for each layout across all scenarios, we get a $T_\text{min}$, which compared to the deterministic case, represents a deviation of -10.69\% in the 12m-layout, -6.32\% in the 20m-layout and -4.41\% in the 28m-layout. In other words, even in the 20m-layout's worst-case scenario, 95\% of all geometries will not undercut the planned systems performance by more than 6.32\%. As shown before, some of them may actually perform better than planned. Only 5\% of the geometries in the worst-case scenario will do even worse than 6.32\% below the planned performance.

In conclusion, the performance of the BHE arrays under consideration is comparatively robust with regards to borehole path uncertainty. Even in the worst of all analyzed scenarios (i.e. 12m-layout with a uniformly distributed inclination of not more than $15\degree$) $T_\text{min}$ only falls short off the respective reference case by -10.69\%. However, the borehole path uncertainty's impact on the BHE array performance is minor to the borehole spacing's influence on the associated thermal interference between BHEs. Considering only the reference cases, the 12m-layout yields a $T_{avg}$ -19.10\% lower than the 20m-layout.

For now, this assessment only applies to the presented case study and other BHE arrays with a similar operation regime and size. To generalize, the underlying effects which led to the observed performance behaviour can be explained using the partially integrated interpolant plotted in Figure \ref{fig:BHE1-9}. The 9 plots show how performance is affected by the individual deviations in the borehole paths for each BHE. In all 9 plots except for BHE 5 (i.e. the central BHE), the performance is negatively affected as the BHE is deviated towards the center of the array and positively affected by being deviated away from the center. This reflects previous considerations and intuition that borehole path deviations potentially reduce the distance between the BHEs increasing thermal interaction among them. In other words, the rock volume for heat extraction is reduced and thereby the arrays performance, as well. Borehole paths pointing away from the center, allow the BHEs to access otherwise untapped regions of soil increasing the overall rock volume for heat extraction. BHE 5 in the center of the array naturally competes with all other BHEs. In the plotted scenario with an allowed vertical angle of up to $6\degree$, BHE 5 cannot be deviated far enough to tap into new soil, which is why we see almost no change in $T_{avg}$. This in fact changes for subsequent scenarios, in which even BHE 5 can be deviated far enough to result in an improvement of performance. For the other 8 border BHEs, there is a trade-off between improvement and deterioration of performance based on their deviation.

This also explains why the 12m-layout is affected so positively by the introduced deviations of the borehole paths. The 12m-layout differs from the 20m-layout only in its BHE spacing, the latter of which is a layout optimized in EED \cite{hellstrom2000earth}. Thus, the basic 12m-layout has to contend with a lot more thermal interaction between the BHEs than recommended. This design flaw is actually alleviated by the introduced deviations in the borehole paths on average, which leads to an increase of the mean performance measure. the drop in mean performance for the 28m-layout can be explained analogously. In the 28m-layout, the BHEs are so far apart, that there is almost no thermal interaction. Therefore, the introduced deviations can only lead to geometries that either maintain the absence of thermal interaction or lead to geometries where BHEs have to compete with each other, leading to a drop in mean performance. The 20m-layout appears to be close the turning point between improvement and deterioration of performance confirming the optimality of its design. While there is a small initial decrease in mean performance for the earlier scenarios ($3\degree$ to $9\degree$), the mean roughly stabilizes for the later scenarios. Instead, whether a geometry leads to improvement or deterioration of performance is largely dependent on the standard deviation.

Generally speaking, there is a trade-off effect between individual BHEs bunching up close to the center of the array leading to a reduced performance and BHEs fanning out from the center improving it. It can be assumed that this effect carries over for larger arrays which either go deeper or include more BHEs, in so far as they are also operated in an extraction regime.

Moreover, arrays with more BHEs should be even more robust with regards to introduced deviations of the borehole paths. This can be explained by the previous observations: As shown in Figure \ref{fig:BHE1-9}, the center BHE is largely negligible in its effect on performance. In scenarios with lesser deviations, its impact can be almost completely ignored, since it cannot escape being in competition with other BHEs. This situation only changes for increasing deviation angles. With BHEs on the border of the array, we see the previously mentioned trade-off. For larger BHE arrays, the ratio of center BHEs to border BHEs increases. By this logic BHE arrays should behave more robustly the larger they are, as on average, center BHEs do not contribute to a deterioration of performance.

Deeper BHE arrays are obviously affected more strongly by deviations in their geometry. A small inclination can lead to dozens of meters of horizontal offset at bottom hole. All of the previously discussed effects leading to improvement and deterioration of performance should be even more pronounced for deeper arrays, since both the bunching up and fanning out of BHEs is more likely to occur for deeper boreholes. Yet, while BHEs arrays can tap much deeper into the soil than considered in this study, the spacing between BHE arrays is typically within the bounds of our assumptions, even for deep BHE arrays. Therefore, borehole paths actually pointing to the center will eventually reach untapped rock on the opposite site of the array and benefit from an enlarged volume for heat extraction. Thus, deep BHE arrays are more likely to profit from deviated borehole paths. Some systems even use directional drilling to deliberately create fanning out BHE arrays to capitalize on this effect \cite{bussmann2015geostar}.

However, deeper BHE arrays are often used as borehole thermal energy storage systems. In contrast to BHE arrays used for heat extraction, borehole thermal energy storage systems, rely on thermal interaction between BHEs and thus on mostly constant BHE spacings. Therefore, borehole path uncertainty needs to be assessed differently in the context of thermal storage, but this issue is beyond the scope of the presented study.

Lastly, the problem of unviable geometries with overlapping borehole paths mentioned in section \ref{sec:modeling} has to be addressed. As stated before, such a case would constitute a critical failure of construction of the BHE array as it would render at least one of the intersecting BHEs inoperable. While borehole path uncertainty in the 12m-layout leads to an increase of the mean $T_{avg}$ suggesting that inaccuracies during drilling can even be a good thing, it has to be pointed out that it is not! The loss of a BHE will most likely overcompensate all gains from preferable borehole path deviations.
Moreover, the presented model, which associates a performance measure with each geometry, cannot sensibly evaluate these cases. For a Monte Carlo sampling method, the cases where such a geometry would occur, could simply be declared inadmissible and ignored. For the stochastic collocation method however, interpolation is performed on the entire support of all random variables. Thus, the small 'patches' on the support which encode unviable geometries cannot be removed. We deal with this problem by slightly correcting any sampled unviable geometries into a viable one. This means that a slightly disturbed interpolation problem is being solved. In practice, the effects of this seem to be negligible and reference solutions calculated with Monte Carlo simulations are still in good agreement. However, an exact convergence between Monte Carlo and stochastic collocation cannot be expected for this disturbed problem.

\section{Conclusion}
\label{sec:conclusion}

In this paper, we present a detailed approach for the assessment of randomly occurring deviations in borehole path geometries and their impact on the performance of BHE arrays. The adaptive, anisotropic stochastic collocation method proves to be a computationally efficient method to perform uncertainty quantification in the context of geothermal applications and provides extensive statistical information on the affected quantities of interest. The comprehensive study showcases the method's application on a BHE array used for heat extraction and demonstrates how the effects of randomly occurring borehole path deviations on the system performance can be quantified.
As a result of the study, heat extraction scenarios in BHE arrays turn out to be surprisingly robust against borehole path uncertainty. Even though random borehole path deviations can be expected during construction, a BHE array will most likely still perform quite similarly to its original planned design.

To our knowledge, this is the first study that investigates the effects of randomly occurring geometric deviations on a BHE array in a rigorous mathematical framework. The presented approach and method provide planners and investors with a useful tool to quantify the risk of geometric uncertainties on the level of individual BHE arrays. The method is not limited to borehole path geometries and can be applied to other parameters in the context of geothermal applications, as well. Aspects like the operation regime, groundwater flow, thermophysical parameters, etc. are typically fraught with uncertainty and could be assessed with the presented method.
In particular, we plan to investigate borehole thermal energy storages, which are operated in a storage-extraction regime. We suspect that the storage coefficient of a borehole thermal energy storage, which is the usual performance measure in this context, could be strongly affected by changes to the geometry and other sources of uncertainty. We also intend to consider different types of geometric deviations, such as nonlinear deviations scaling with bore depth, as this constitutes a more realistic behaviour.

\section*{Acknowledgements}
\label{sec:acknowledgements}

This Research is supported by the Darmstadt Graduate School of Excellence Energy Science and Engineering (GSC 1070) within the framework of the Excellence Initiative by the German federal and state governments.

\bibliographystyle{plain}
\bibliography{uqgeo}

\end{document}